\documentclass[twocolumn]{aastex631}

\usepackage{graphicx,grffile,ulem,url,threeparttable,xcolor,soul,booktabs,natbib,amsmath,mathrsfs,makecell}
\usepackage{txfonts,rotating,longtable, needspace,subfigure,multirow,array,hyperref}
\usepackage{orcidlink}

\shorttitle{Hot Intragroup Medium in Poor Groups}
\shortauthors{Li et al.}

\graphicspath{{./}{figures/}}

\begin{document}


\title{Robust detection of hot intragroup medium in optically selected, poor galaxy groups by eROSITA}


\author[0009-0000-5534-1935]{Dawei Li\orcidlink{0009-0000-5534-1935}}
\affiliation{Department of Astronomy, Xiamen University, Xiamen, Fujian 361005, People's Republic of China}

\author[0000-0002-2853-3808]{Taotao Fang\orcidlink{0000-0002-2853-3808}}
\affiliation{Department of Astronomy, Xiamen University, Xiamen, Fujian 361005, People's Republic of China}

\author[0000-0003-0628-5118]{Chong Ge\orcidlink{0000-0003-0628-5118}}
\affiliation{Department of Astronomy, Xiamen University, Xiamen, Fujian 361005, People's Republic of China}

\author[0000-0002-2941-6734]{Teng Liu\orcidlink{0000-0002-2941-6734}}
\affiliation{Department of Astronomy, University of Science and Technology of China, Hefei 230026, People's Republic of China}
\affiliation{School of Astronomy and Space Science, University of Science and Technology of China, Hefei 230026, People's Republic of China}

\author[0000-0002-7875-9733]{Lin He\orcidlink{0000-0002-7875-9733}}
\affiliation{School of Astronomy and Space Science, Nanjing University, Nanjing 210023, People's Republic of China}
\affiliation{Key Laboratory of Modern Astronomy and Astrophysics (Nanjing University), Ministry of Education, Nanjing 210023, People's Republic of China}

\author[0000-0003-0355-6437]{Zhiyuan Li\orcidlink{0000-0003-0355-6437}}
\affiliation{School of Astronomy and Space Science, Nanjing University, Nanjing 210023, People's Republic of China}
\affiliation{Key Laboratory of Modern Astronomy and Astrophysics (Nanjing University), Ministry of Education, Nanjing 210023, People's Republic of China}

\author[0000-0002-6896-1364]{Fabrizio Nicastro\orcidlink{0000-0002-6896-1364}}
\affiliation{Department of Astronomy, Xiamen University, Xiamen, Fujian 361005, People's Republic of China}
\affiliation{Istituto Nazionale di Astrofisica (INAF)—Osservatorio Astronomico di Roma Via Frascati, 33 I-00078 Monte Porzio Catone (RM), Italy}

\author[0000-0003-3997-4606]{Xiaohu Yang\orcidlink{0000-0003-3997-4606}}
\affiliation{Department of Astronomy, School of Physics and Astronomy, and Shanghai Key Laboratory for Particle Physics and Cosmology, Shanghai Jiao Tong University, Shanghai 200240, People's Republic of China}
\affiliation{Tsung-Dao Lee Institute, and Key Laboratory for Particle Physics, Astrophysics and Cosmology, Ministry of Education, Shanghai Jiao Tong University, Shanghai 200240, People's Republic of China}

\author[0000-0003-4832-9422]{Xiaoxia Zhang\orcidlink{0000-0003-4832-9422}}
\affiliation{Department of Astronomy, Xiamen University, Xiamen, Fujian 361005, People's Republic of China}

\author[0000-0002-5632-9345]{Yun-Liang Zheng\orcidlink{0000-0002-5632-9345}}
\affiliation{Department of Astronomy, School of Physics and Astronomy, and Shanghai Key Laboratory for Particle Physics and Cosmology, Shanghai Jiao Tong University, Shanghai 200240, People's Republic of China}

\correspondingauthor{Taotao Fang, Chong Ge}
\email{fangt@xmu.edu.cn, chongge@xmu.edu.cn}

\begin{abstract}
Over the last several decades, extensive research has been conducted on the baryon cycles within cosmic structures, encompassing a broad mass range from dwarf galaxies to galaxy clusters. However, a notable gap in understanding the cosmic baryon cycle is the poor galaxy groups with halo masses around $10^{13}\ M_{\odot}$ \citep[e.g.,][]{McGaugh_2010_ApJ}. Poor galaxy groups, like our own Local Group, are prevalent throughout the universe, yet robust detection of their hot, X-ray emitting intragroup medium (IGrM) has remained elusive. The presence of this hot IGrM is crucial for addressing the long-standing ``missing baryons'' problem. Previous ROSAT-based studies were limited by a small number of X-ray bright samples, thus restricting the scope of their findings. Here we show a robust detection of this hot IGrM in a large, optically selected poor groups sample, based on the stacked X-ray images from the eROSITA Final Equatorial Depth Survey. These groups are identified in DESI LS with a mass range of log($M_\mathrm{halo}/h^{-1}M_{\odot}$) = 11.5--13.5 and a redshift range of z = 0.1--0.5. Additionally, our results indicate that despite its presence in virtually groups at all sizes, this gas component is still not sufficient to recover the universal baryon fraction, and hence the ``missing baryons" problem still persists in poor galaxy groups.
\end{abstract}

\section{Introduction} \label{sec:introduction}
Poor galaxy groups, characterized by their small number of member galaxies, usually exhibit lower halo mass and virial temperature when compared with typical galaxy groups and clusters. The Local Group, home to the Milky Way and Andromeda, serves as a notable example of such a system. These poor groups, hosting a significant amount of galaxies, play a vital role in cosmic structure formation and evolution \citep{Tully_87_APJ, Eke_04_MNRAS}. Due to the prevalence of these systems, several studies have suggested that the substantial ``missing baryons'' in the nearby universe are likely situated in the intragroup medium (IGrM) consisting of hot X-ray-emitting gas (T$>$$10^6$ K) within poor groups \citep{Fukugita_1998_ApJ, Cen_2001_ApJ...552..473D, Cen_2006_ApJ...650..560C,Shull_2012_ApJ...759...23S, Tuominen_2021_A&A...646A.156T}. Despite this, robust detections of the hot IGrM in poor groups have been scarce in the past \citep{Mulchaey_00_ARAA}. Early X-ray observations by Einstein and ROSAT identified diffuse X-ray emission in some poor groups \citep[e.g.,][]{Mulchaey_98_APJ, Helsdon_2000_MNRAS}, suggesting the presence of an intragroup medium (IGrM). There have also been in-depth studies of poor groups using the modern X-ray observatories Chandra and XMM-Newton \citep[e.g.,][]{Helsdon_2005_ApJ, Rasmussen_2006_MNRAS}, which have further investigated the properties of the IGrM in poor groups. However, these studies are limited by small sample sizes and X-ray selection effects, thus restricting the generalizability of their findings. Consequently, the characterization of the diffuse IGrM within poor groups remains an open issue, warranting further investigation and exploration.

The newly launched eROSITA telescope's advancements have made the study of the IGrM feasible, with its 30-50 times greater sensitivity than ROSAT and a two-fold improvement in angular resolution \citep{Merloni_20_NA}. The eROSITA Final Equatorial Depth Survey (eFEDS) covers a contiguous 140 deg$^2$ field with nearly uniform exposure of approximately 2.5 ks, expected to be representative of the end-of-survey all-sky exposure \citep{Predehl_21_A&A, Brunner_2022_A&A}. Leveraging eROSITA's superior sensitivity and spatial resolution, we used the stacking technique on eFEDS data to detect the hot IGrM within poor groups. Previous stacking studies using eROSITA data have successfully examined the circumgalactic medium (CGM) of galaxies and the IGrM of galaxy groups \citep[e.g.,][]{Comparat_22_A&A, Chadayammuri_22_APJL, Popesso_2024_MNRAS, Zheng_23_MNRAS, Zhang_2024arXiv240117308Z}. In particular, \cite{Zhang_2024arXiv240117308Z} stacked the first four eROSITA all-sky survey data on a large, optically selected galaxy sample to study the CGM of Milky Way-sized galaxies. The halo mass range of their sample partially overlaps with that of this work. However, their sample selection and focus differ from ours, which aims to investigate the IGrM within poor groups, rather than individual galaxies. Additionally, we constructed a control sample to verify that the CGM is not the significant component in the stacked signal, as discussed in Section \ref{sec:discuss}.

The organization of this paper is as follows: In Section \ref{sec:methonds}, we provide a description of the data employed in this study, as well as the primary stacking methods. Section \ref{sec:results} outlines the main findings regarding X-ray luminosity, gas mass, and gas fraction ($M_\mathrm{gas}$/$M_\mathrm{halo}$). In Section \ref{sec:discuss}, we compare our results with previous studies and discuss potential sources of contamination. Finally, we summarize our findings in Section \ref{sec:summary}. Throughout this paper, we adopt a $\Lambda$CDM cosmology consistent with the Planck 2018 results \citep{Planck18_20_A&A}. If not specified otherwise, the X-ray
luminosity $L_\mathrm{X}$ is given in the range of 0.5--2.0 keV.

\section{Samples and Data Analysis} \label{sec:methonds}

\subsection{Samples}
To eliminate X-ray selection bias, we used a large, optically selected group catalog from \cite{Yang_2021_APJ}, based on the DESI Legacy Imaging Surveys \citep[]{DESI_2019_AJ}. \cite{Yang_2021_APJ} expanded their halo-based group finder \citep{Yang_2005_MNRAS, Yang_2007_ApJ} to select galaxy groups using data from the DESI LS DR8, incorporating both photometric and spectroscopic redshift information, resulting in a large and comprehensive galaxy group sample. To ensure the highest possible accuracy of redshift information, a small fraction of the redshifts were replaced with the latest available spectroscopic measurements. In our study, we used this updated catalog and specifically selected groups with spectroscopic redshifts. We then cross-matched our group catalog with the eFEDS survey and applied several criteria for the selection of poor groups. Since \cite{Zabludoff_1998_ApJ} defined a poor group as an apparent system with fewer than 5 bright galaxies, we restricted our selection to groups with 2-4 member galaxies. Considering the moderate angular resolution of eROSITA, we imposed an upper redshift limit of 0.5. Due to the scarcity of poor groups below redshift 0.1, we narrowed down the redshift range to 0.1-0.5. Specifically, the cross-matching revealed that there are only 403 groups below redshift 0.1, with more than 100 groups only in the lowest mass range, making these samples unsuitable for our analysis. This scarcity is primarily caused by the volume-limited nature of the survey, which reduces the observed spatial volume and thus limits the number of detectable groups \citep[refer to][]{Yang_2021_APJ}. In the end, we selected a total sample of 25,524 poor groups for subsequent stacking analysis. To study the redshift evolution, we also divided these final samples into four redshifts ranges: 0.1 $<$ $z$ $<$ 0.2, 0.2 $<$ $z$ $<$ 0.3, 0.3 $<$ $z$ $<$ 0.4 and 0.4 $<$ $z$ $<$ 0.5. Furthermore, we categorized them within each redshift bin into four halo mass ranges: log($M_\mathrm{halo}$/$h^{-1}M_\mathrm{\odot}$) = [11.5, 12], [12, 12.5], [12.5, 13], [13, 13.5]. The number of poor groups within each mass and redshift range is shown in Table \ref{tab:table1}.

Recently, the eROSITA team released the first all-sky survey data  \citep[eRASS1,][]{Merloni_2024_A&A}. We found that the overlapped area between eRASS1 and our group catalog is around 20 times greater than that between eFEDS and the group catalog. However, the average exposure time of most areas in eRASS1 is only about 1/10 of eFEDS, leading to only a slight improvement in the overall signal-to-noise ratio (S/N) when using eRASS1 data compared to eFEDS data. Therefore, in this work, we used only eFEDS data for the X-ray analysis. Further investigation using eRASS1 data lies beyond the scope of this study and will be considered in future work.

\begin{table*}[htbp]
\small
\centering
\caption{Summary of Sample Characteristics, Fitted Parameters, and Derived Results.}
\label{tab:table1}
\begin{tabular*}{\textwidth}{@{\extracolsep{\fill}}ccccccccccc}
\midrule \midrule
Redshift & Group Mass & Number & S/N & Counts & $R_c$ & $\beta$ & $L_\mathrm{0.5-2.0 keV}$ & $M_\mathrm{gas}$ & $f_\mathrm{baryon}$ & $f_\mathrm{cont}$ \\
(1) & (2) & (3) & (4) & (5) & (6) & (7) & (8) & (9) & (10) & (11)
\\
\midrule
\multirow{4}{*}{0.1--0.2} & 11.5--12.0 & 1415 & 3.96 (136) & 14460 & 4.59$^{+10.31}_{-3.28}$ & 0.34$^{+0.02}_{-0.02}$ & 0.16$^{+2.44}_{-0.15}$ & 0.33$^{+0.97}_{-0.25}$ & 0.12$^{+0.21}_{-0.05}$ & 0.53$^{+1}_{-0.45}$\\
& 12.0--12.5 & 1830 & 6.38 (164) & 27314 & 4.05$^{+10.64}_{-3.09}$ & 0.32$^{+0.02}_{-0.02}$ & 0.45$^{+7.07}_{-0.42}$ & 0.65$^{+1.96}_{-0.50}$ & 0.09$^{+0.15}_{-0.04}$ & 0.33$^{+1}_{-0.30}$\\
& 12.5--13.0 & 1100 & 8.89 (168) & 17158 & 5.72$^{+12.48}_{-3.99}$ & 0.34$^{+0.01}_{-0.01}$ & 1.50$^{+16.25}_{-1.38}$ & 1.18$^{+2.86}_{-0.85}$ & 0.07$^{+0.07}_{-0.02}$ & 0.15$^{+1}_{-0.13}$\\
& 13.0--13.5 & 225 & 6.35 (449) & 24438 & 32.96$^{+32.41}_{-23.76}$ & 0.41$^{+0.07}_{-0.04}$ & 2.63$^{+22.39}_{-2.53}$ & 2.18$^{+4.60}_{-1.84}$ & 0.05$^{+0.04}_{-0.01}$ & 0.14$^{+1}_{-0.12}$\\
\midrule
\multirow{4}{*}{0.2--0.3} & 11.5--12.0 & 943 & 0.87 &  \dots  & \dots & \dots & \dots & \dots & \dots & \dots\\
& 12.0--12.5 & 3893 & 5.93 (110) & 14900 & 3.45$^{+10.64}_{-3.09}$ & 0.32$^{+0.02}_{-0.02}$ & 0.73$^{+18.91}_{-0.71}$ & 0.69$^{+2.84}_{-0.57}$ & 0.09$^{+0.21}_{-0.04}$ & 0.24$^{+1}_{-0.22}$\\
& 12.5--13.0 & 2783 & 11.21 (128) & 15470 & 7.69$^{+14.58}_{-5.19}$ & 0.37$^{+0.01}_{-0.01}$ & 2.36$^{+31.92}_{-2.22}$ & 1.21$^{+3.38}_{-0.91}$ & 0.07$^{+0.09}_{-0.02}$ & 0.12$^{+0.32}_{-0.11}$\\
& 13.0--13.5 & 582 & 11.57 (313) & 17314 & 59.12$^{+24.40}_{-25.10}$ & 0.41$^{+0.03}_{-0.03}$ & 7.38$^{+11.28}_{-5.01}$ & 3.10$^{+1.78}_{-1.39}$ & 0.06$^{+0.01}_{-0.01}$ & 0.05$^{+0.12}_{-0.03}$\\
\midrule
\multirow{4}{*}{0.3--0.4} & 11.5--12.0 & 49 & 0.82 & \dots  & \dots & \dots & \dots & \dots & \dots & \dots\\
& 12.0--12.5 & 2342 & 4.41 (133) & 9046 & 5.22$^{+21.93}_{-4.21}$ & 0.37$^{+0.03}_{-0.03}$ & 1.42$^{+62.37}_{-1.28}$ & 0.85$^{+4.81}_{-0.75}$ & 0.10$^{+0.30}_{-0.05}$ & 0.20$^{+1}_{-0.14}$\\
& 12.5--13.0 & 4103 & 10.61 (155) & 22179 & 8.13$^{+18.38}_{-6.33}$ & 0.36$^{+0.01}_{-0.01}$ & 3.54$^{+72.80}_{-3.39}$ & 1.24$^{+4.61}_{-0.99}$ & 0.07$^{+0.11}_{-0.02}$ & 0.09$^{+0.33}_{-0.08}$\\
& 13.0--13.5 & 1243 & 12.04 (221) & 13302 & 88.52$^{+8.41}_{-14.47}$ & 0.49$^{+0.03}_{-0.03}$ & 8.52$^{+4.47}_{-3.62}$ & 2.86$^{+0.66}_{-0.73}$ & 0.06$^{+0.01}_{-0.01}$ & 0.06$^{+0.04}_{-0.02}$\\
\midrule
\multirow{4}{*}{0.4--0.5} & 11.5--12.0 & 3 & -0.18 & \dots  & \dots & \dots & \dots & \dots & \dots & \dots\\
& 12.0--12.5 & 383 & 0.36 & \dots  & \dots & \dots & \dots & \dots & \dots & \dots\\
& 12.5--13.0 & 2992 & 8.65 (201) & 20062 & 11.89$^{+26.48}_{-10.01}$ & 0.31$^{+0.02}_{-0.01}$ & 4.86$^{+63.89}_{-4.57}$ & 1.25$^{+3.33}_{-0.94}$ & 0.06$^{+0.07}_{-0.02}$ & 0.07$^{+1}_{-0.06}$\\
& 13.0--13.5 & 1638 & 7.12 (301) & 24368 & 62.58$^{+24.68}_{-36.12}$ & 0.47$^{+0.06}_{-0.05}$ & 7.68$^{+22.83}_{-6.59}$ & 2.25$^{+2.20}_{-1.51}$ & 0.05$^{+0.02}_{-0.01}$ & 0.08$^{+0.50}_{-0.05}$\\
\midrule
\end{tabular*}
\textbf{Note.} Column (1): redshift range of each subsample. Column (2): logarithm of the halo mass (unit: $h^{-1}M_\mathrm{\odot}$) range of each subsample. Columns (3): number of groups within each subsample. Column (4): the maximum S/N of the hot IGrM emission and the corresponding detection radius (unit: kpc). Column (5): total counts within the source region corresponding to the maximum S/N. Column (6-7): best-fit values and 1$\sigma$ uncertainties of beta model parameters. The unit of $R_c$ is kpc. Column (8): estimated X-ray luminosity within $r_{500}$ (unit: $10^{41}$erg/s). Column (9): estimated gas mass within $r_{500}$ (unit: $10^{11} M\mathrm{_\odot}$). Column (10): estimated baryon fraction within $r_{500}$. Column (11): estimated contamination fraction. If the upper limit of the error bar exceeds 1, it is capped at 1.
\end{table*}

\subsection{eROSITA Data Analysis}

We adopted the exposure-corrected eFEDS map within the energy range of 0.2--2.3 keV. Images were generated by employing the eFEDS event files and the early version of the eROSITA Science Analysis Software System (\texttt{eSASS}), which is available in the Early Data Release website\footnote{\href{https://erosita.mpe.mpg.de/edr/DataAnalysis/}{https://erosita.mpe.mpg.de/edr/DataAnalysis/}}. We converted the event file into an image using the \texttt{evtool} command and produced a corresponding exposure map using the \texttt{expmap} command. The average exposure time across most of the eFEDS field is approximately 1.2 ks after correcting for vignetting.

We employed the catalog generated by \cite{Brunner_2022_A&A} and \cite{Liu_Ang_22_A&A} to mask the previously detected sources in the X-ray images. The mask radius was set to 30 arcseconds for point sources. For extended sources, we adopted the spatial extent radius provided in \cite{Liu_Ang_22_A&A}'s catalog. To ensure accurate background subtraction, in addition we visually checked for any obvious excess emission regions and excluded them. We applied the same mask procedures to the exposure map. We obtained X-ray images and exposure maps with the detected sources subtracted.

\begin{figure*}[htbp]
    \centering
    \includegraphics[width=0.8\textwidth, keepaspectratio]{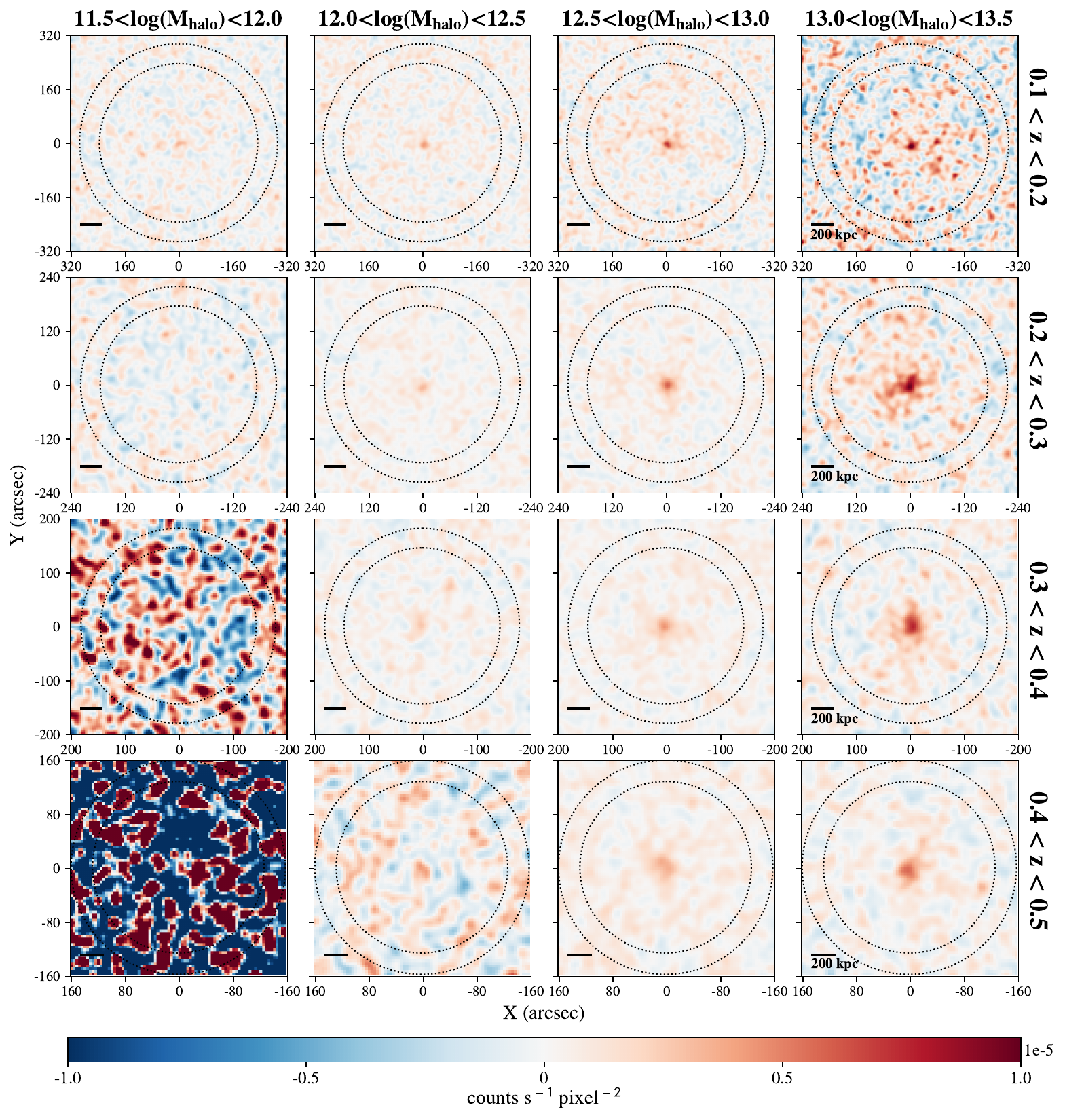}
    \caption{Background-subtracted surface brightness images of all subsamples in the 0.2--2.3 keV band. The annulus shows the region used to compute the background. Emission is seen in most subsamples but is more extended and brighter in the high-mass groups subsamples. The scale bar corresponds to 200 kpc in each panel. The unit of halo mass ($M_\mathrm{halo}$) is: $h^{-1}M_\mathrm{\odot}$.}
    \label{fig:image}
\end{figure*}

The centers of the groups were defined as luminosity-weighted centers based on the DESI LS data \citep{Yang_2021_APJ}. For the stacking analysis, we first extracted individual galaxy group images with fixed sizes from the eFEDS X-ray image and exposure map. These fixed sizes were chosen to ensure that the extracted regions exceeded the virial radius of the groups, providing sufficient coverage for all systems. We then rescaled the images by adopting the angular size of the group with the highest redshift in each subsample as the standard size. This rescaling adjusted both the pixel positions and pixel counts of the images through interpolation. As the precision of interpolation is limited, we chose the highest redshift for the standard size to minimize uncertainties, since downscaling introduces smaller errors compared to upscaling. Finally, we stacked the rescaled X-ray images of the groups across all subsamples.

\cite{Dai_07_APJ} and \cite{Anderson_2015_MNRAS.449.3806A} employed similar X-ray stacking methods, but their approaches differ from ours in several aspects. In \cite{Dai_07_APJ}, photons were weighted during image rescaling by the square of the ratio of the luminosity distance of the group to that of the reference redshift. However, this method affects not only the photons from the group but also the background photons, potentially introducing uncertainties in the analysis of faint sources due to the critical role of background levels. In contrast, \cite{Anderson_2015_MNRAS.449.3806A} performed stacking in physical space without weighting photon counts, which introduced biases toward the nearest sources.

In our method, we rescaled images without applying additional weighting to individual photon counts. Instead, interpolation was used to adjust pixel counts, ensuring that the background level and surface brightness of the groups remain unchanged. Since the images are already weighted by the square of the angular size as a whole, this approach minimizes biases in the photon distribution toward the nearest groups.

To account for the background, we defined an annulus ranging from 800--1000 kpc around the group center to compute the average count rate as its background value. The median background value of all groups within each subsample was used as the corresponding background. Additionally, we computed the background in three annuli: 800--1000 kpc, 600--800 kpc, and 1000--1200 kpc, finding no significant differences among these values (all approximately 2.6 × $10^{-5}$ cts/s/pixel$^{2}$), consistent with the background values reported in other studies \citep[e.g.,][]{Zheng_23_MNRAS}.

\begin{figure*}[htbp]
    \centering
    \includegraphics[width=0.8\textwidth, keepaspectratio]{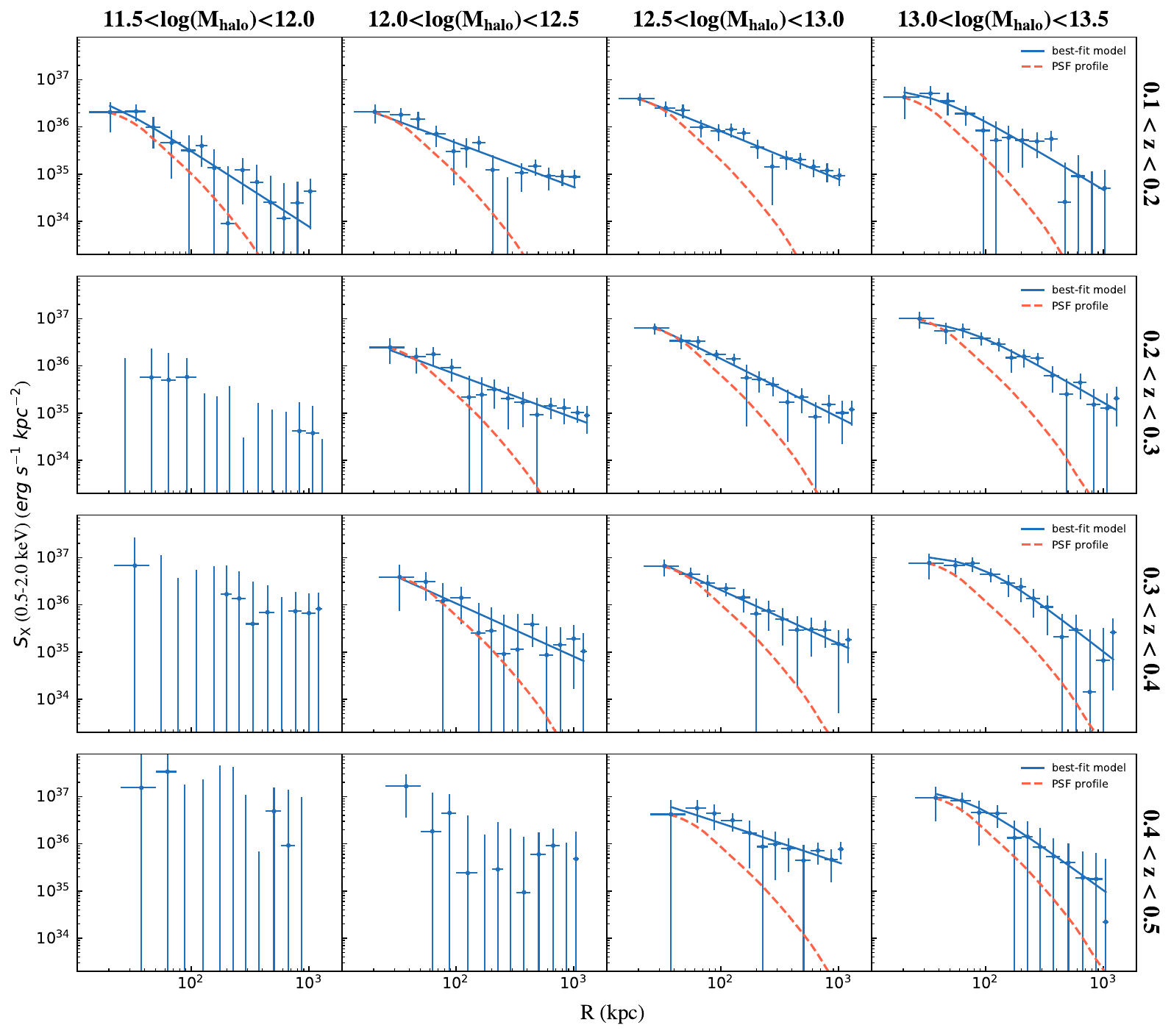}
    \caption{The surface brightness profiles of subsamples. Blue crosses show the background-subtracted profile. The blue solid lines represent the best-fit $\beta$-model for each subsample, except for the four undetected subsamples. The data error bars correspond to the quadratic sum of photon Poisson errors and the stacking uncertainty. The orange dotted lines represent the PSF profile of eROSITA. The unit of halo mass ($M_\mathrm{halo}$) is: $h^{-1}M_\mathrm{\odot}$.}
    \label{fig:profile}
\end{figure*}

\section{Results} \label{sec:results}

\subsection{S/N Estimation of Stacked Signal}

After obtaining the stacked images (shown in Fig. \ref{fig:image}), we calculated the significance of the central excess emission from the hot IGrM based on:
\begin{equation}
S/N = \frac{N_\mathrm{s}-N_\mathrm{b}}{\sqrt{N_\mathrm{s}}},
\end{equation}
where $N_\mathrm{s}$ represents the total counts of source regions within the radius where the highest S/N value is achieved, and $N_\mathrm{b}$ represents the corresponding background counts. These background counts were obtained by using the derived median background values in each subsample and multiplying them by the total exposure time within the corresponding source region. The best S/N estimates of the hot IGrM emission and source counts are presented in Table \ref{tab:table1}.

Twelve out of sixteen subsamples exhibit strong excess emission beyond the background with significances ranging from 3.9$\sigma$ to 12$\sigma$. The three non-detections in the lowest halo mass range at redshifts higher than 0.2, along with the one non-detection in the second lowest halo mass range at the highest redshift, are likely due to the diminishing number of poor groups identified in DESI LS and the limited survey depth of eFEDS.

\begin{figure*}[tb]
    \centering
    \includegraphics[width=0.75\textwidth, keepaspectratio]{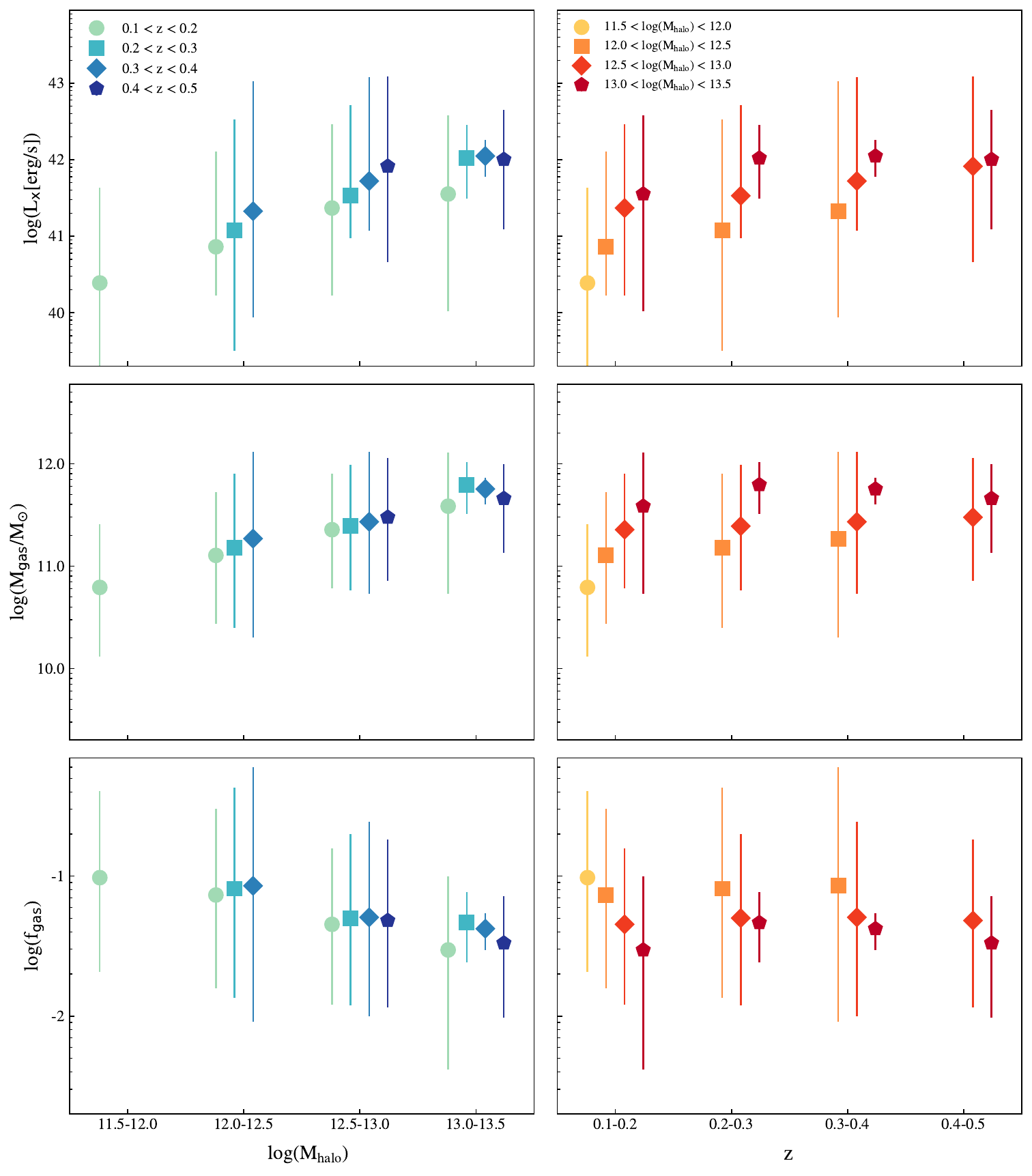}
    \caption{From top to bottom, the three rows represent the results for luminosity, gas mass, and gas fraction within $r_{180}$, respectively. The first column uses halo mass as the X-axis, while the second column uses redshift as the X-axis. The error bars correspond to the 1$\sigma$ errors of the best-fit parameters. The unit of halo mass ($M_\mathrm{halo}$) is: $h^{-1}M_\mathrm{\odot}$.}
    \label{fig:results}
\end{figure*}

\begin{figure*}[htbp]
    \centering
    \includegraphics[width=0.9\textwidth, keepaspectratio]{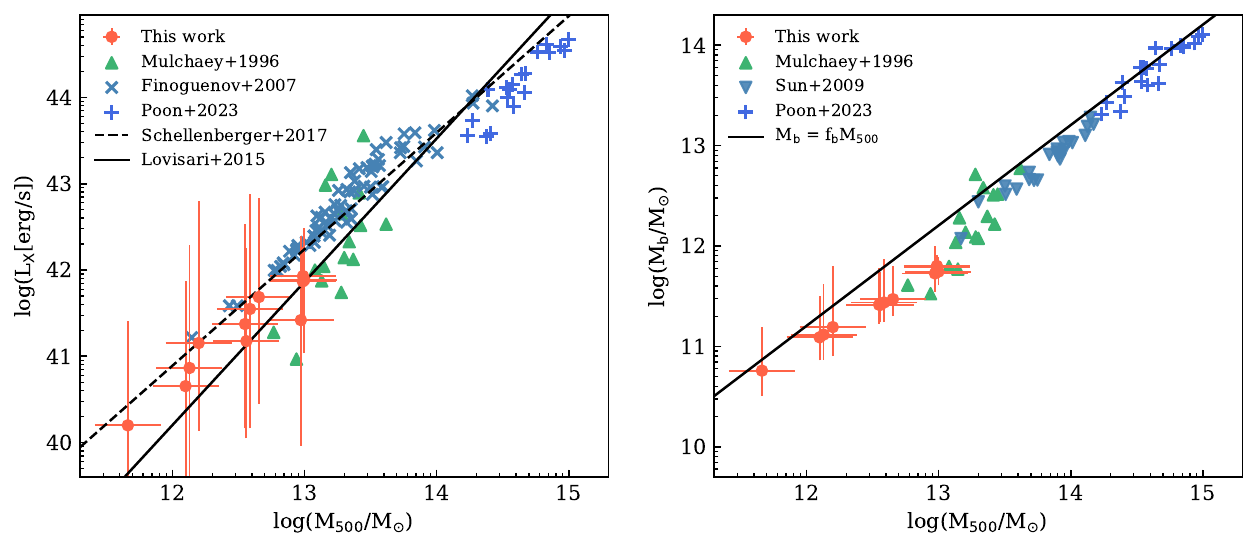}
    \caption{Comparison of luminosities and baryon masses within $r_{500}$ between this work (red points) and previous studies, including poor galaxy groups (green) and rich galaxy groups and clusters (blue). In the left panel, the solid line represents the $L_\mathrm{X}$-$M_\mathrm{500}$ relation for galaxy groups from \cite{Lovisari_2015_A&A...573A.118L}, while the dashed line represents the $L_\mathrm{X}$-$M_\mathrm{500}$ relation for galaxy clusters from \cite{Schellenberger_2017_MNRAS}. The solid line of right panel displays the predicted $M_\mathrm{b}$ at corresponding $M_\mathrm{500}$ based on the $\Lambda$CDM model.}
    \label{fig:comparison}
\end{figure*}

\subsection{Surface Brightness Profiles}

To assess the extent of the excess emission, we generated surface brightness profiles (shown in Fig. \ref{fig:profile}) of the stacked images and fitted them with the standard $\beta$-model. We generated radial profiles by creating multiple annular regions extending outward from the center of the stacked images, calculating the average counts rate within each annular region. To ensure an adequate counts, we set each annular region to have a ratio of R$_{out}$/R$_{in}$ = 1.3. 

The uncertainties of the profiles arise from two sources: Poisson errors and uncertainties from the stacking process. We applied the Jackknife re-sampling method \citep[][]{Andrae_2010_arXiv1009.2755A, McIntosh_2016_arXiv160600497M} to estimate the uncertainty due to the stacking process. For each subsample, we randomly stacked 90\% of the poor groups and generated the corresponding profile, repeating this process 50 times. The standard deviation of the 50 trials was taken as the stacking uncertainty. Finally, we calculated the total uncertainty of the profiles as the quadratic sum of the Poisson error and the stacking uncertainty.

Some subsamples exhibit high S/N radial profiles even extending up to the virial radius, suggesting a potentially large extent of the hot IGrM. We also plot the eROSITA point-spread-function (PSF) profile in Fig. \ref{fig:profile}, from which it can be seen that the stacked signal is much more extended than the PSF.

We then employed the standard $\beta$-model \citep{Cavaliere_76_A&A} to fit the profiles:
\begin{equation}
I(r) = I_0 \left(1 + \frac{r^2}{{r_\mathrm{c}}^2}\right)^{1/2-3\beta},
\end{equation}
where $I_0$ represents the central surface brightness, $r_c$ denotes the core radius, and $\beta$ corresponds to the slope of the radial profile. We performed a Markov Chain Monte Carlo (MCMC) analysis to obtain the best-fitting parameters along with their 1$\sigma$ ranges for all subsamples. The results are summarized in Table \ref{tab:table1}.

Dark matter halos are commonly defined as having an overdensity of 180 times the background density of the universe based on the spherical overdensity algorithm \citep[e.g.,][]{Cole_1988_MNRAS, Watson_2013_MNRAS}{}{}. We can determine the comoving halo radius $r_{180}$ via this equation \citep{Yang_2021_APJ}:
\begin{equation}
r_{180} =  0.781h^{-1}\mathrm{Mpc} \left(\frac{M_L}{\Omega_m10^{14}h^{-1}M_\mathrm{\odot}}\right)^{1/3}.
\end{equation}
Subsequently, we integrated the best-fit $\beta$-model up to $r_\mathrm{180}$ to obtain the total background-subtracted counts. To calculate the energy conversion factors (ECFs), we used APEC models set with four distinct temperature and metallicity parameters for different halo mass bins. Temperature and metallicity of each mass range were adopted from the $T-M$ relation in \cite{Sun_09_APJ} and the $Z-M$ relation in \cite{Truong_2019_MNRAS}, respectively. We set the column density parameters at $N_\mathrm{H} = 3 \times 10^{20}$ cm$^{-2}$, as other studies \citep[e.g.,][]{Chadayammuri_22_APJL}. Using the derived ECFs, we converted the counts to rest-frame luminosity in the energy range of 0.5 to 2.0 keV for each subsample. 

We estimated the mass of the hot IGrM within $r_\mathrm{180}$ following the approach described in \cite{Ge_16_MNRAS}. We assumed a spherically symmetric distribution and adopted the deprojected hydrogen density profile of the beta model as a function of the physical off-center radius ($r$) of a group \citep{Sarazin_88_book},
\begin{equation}
    n_\mathrm{H} = n_\mathrm{0}\left(1+\frac{r^2}{{r_\mathrm{c}}^2}\right)^{-\frac{3}{2}\beta},
\end{equation}
where $n_\mathrm{0}$ can be expressed as:
\begin{equation}
n_\mathrm{0} = \frac{180}{\mathrm{\pi}}\sqrt{\vphantom{}\frac{10^{14}4\sqrt{\mathrm{\pi}}I_\mathrm{0}\Gamma(3\beta)}{(\frac{n_\mathrm{e}}{n_\mathrm{H}})\frac{\mathrm{CR}}{\eta}r_\mathrm{c}\Gamma(3\beta-1/2)}}.
\end{equation}
The parameter values of $I_\mathrm{0}$, $r_\mathrm{c}$, and $\beta$ were obtained from the $\beta$-model fitting. We assumed $n_\mathrm{e} = 1.2n_\mathrm{H}$ and used the same APEC models as earlier to determine the $\frac{\mathrm{CR}}{\eta}$ (count rate/normalization). We refer to \cite{Ge_16_MNRAS} for details.

In Figure \ref{fig:results}, we present the resulting luminosity, gas mass, and gas fraction ($M_\mathrm{gas}$/$M_\mathrm{halo}$) for all subsamples within the $r_\mathrm{180}$ range. The error bars were calculated based on the 1$\sigma$ ranges of the fitting parameters. Since the S/N of the subsamples, especially in low-mass subsamples, is relatively low, the fitting results have larger uncertainties, resulting in larger error bars. The estimated luminosity of all subsamples falls within the range of $10^{40}$ to $10^{42}$ erg/s, while the gas mass estimates range from $10^{10.5}$ to $10^{12.0}$ $M_\mathrm{\odot}$. The gas fraction estimates for all subsamples are around 6\%. There is a positive correlation between luminosity and halo mass, as well as between gas mass and halo mass. We did not observe any significant redshift evolution effects.

\begin{figure*}[htbp]
    \centering
    \includegraphics[width=0.9\textwidth, keepaspectratio]{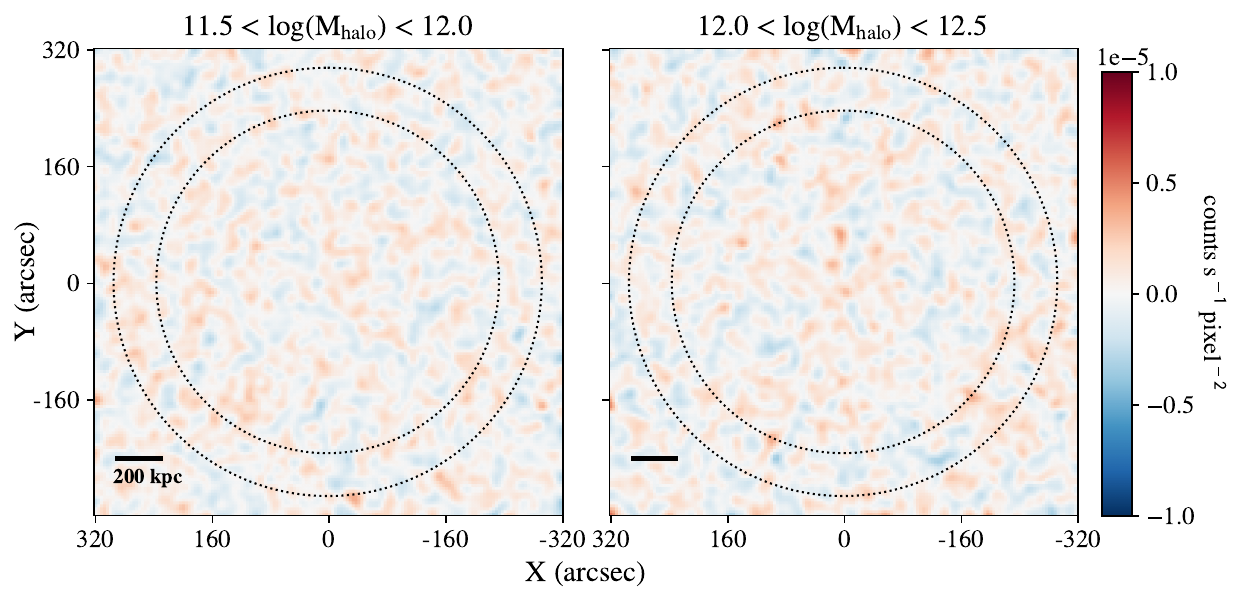}
    \caption{Background-subtracted surface brightness images for two control subsamples in the 0.2--2.3 keV band. Both images correspond to the redshift range of 0.1--0.2, with the left panel corresponding to the halo mass range of 11.5--12.0 and the right panel corresponding to the halo mass range of 12.0--12.5. The annulus represents the region used for background computation. The unit of halo mass ($M_\mathrm{halo}$) is: $h^{-1}M_\mathrm{\odot}$.}
    \label{fig:control_image}
\end{figure*}

\section{Discussion} \label{sec:discuss}

\subsection{Comparison with Previous Studies}

\subsubsection{The $\beta$ Parameter}
The parameter $\beta$ is crucial, as it represents the slope of surface brightness profiles and also reflects the concentration of gas distribution within the poor groups. In our study, the best-fit values of $\beta$ (as listed in Table \ref{tab:table1}) range from 0.32 to 0.49, which is slightly lower than values reported in prior studies on galaxy groups \citep[around 0.5,][]{Helsdon_2000_MNRAS} and clusters \citep[around 0.65,][]{Mohr_1999_ApJ, Sanderson_2003_MNRAS}. This difference aligns with our expectation that these small groups tend to exhibit flatter surface brightness profiles compared to massive systems, likely because they are still in an ongoing formation stage \citep[e.g.,][]{Sanderson_2003_MNRAS, Sun_2012_NJPH}. Additionally, our analysis shows a slight increasing trend in $\beta$ values as halo mass increases within the same redshift. We did not observe any redshift dependence in $\beta$ values.

\subsubsection{Comparison with Stacking-Based Studies}
\cite{Popesso_2024_MNRAS} recently constructed an optically selected GAMA groups and clusters sample undetected by eROSITA, stacking their eFEDS images to study their properties. Unlike our work, their sample was restricted to systems with a richness of $\geq 5$, corresponding to a higher halo mass range of $\sim10^{13}-5\times10^{14}M_\odot$. Their luminosity estimates in the 0.5--2.0 keV band range from $10^{41}$ to $10^{43}$ erg/s, which are higher than our estimates, likely due to the higher halo mass range of their sample. Additionally, their gas fraction estimates, approximately 6\% within $r_{200}$, are comparable to the mean gas fraction we derived across all subsamples within $r_{180}$.

\cite{Zheng_23_MNRAS} used the same group catalog as this work and also stacked eFEDS images to investigate the $L_\mathrm{X}$-$M$ relation of galaxy groups and clusters. Their sample spans a broader halo mass range of $10^{11}-10^{15}M_\odot$, but they did not distinguish between galaxy clusters, galaxy groups, and isolated galaxies within this range. This lack of separation resulted in their luminosity estimates being a mixture of these three types of systems. In contrast, our study specifically focuses on poor galaxy groups. Despite this difference, we found that our luminosity estimates, ranging from $10^{40}$ to $10^{42}$ erg/s, are consistent with their results within the halo mass range considered in this work.

Earlier stacking analyses based on ROSAT data by \cite{Dai_07_APJ} and \cite{Anderson_2015_MNRAS.449.3806A} also utilized optically-selected samples to stack X-ray images. However, the aims and properties of their samples differ from ours. \cite{Dai_07_APJ} focused on galaxy clusters with higher halo masses than those in our sample, leading to higher luminosity estimates. Nonetheless, they also reported a declining trend in the $\beta$ parameter with decreasing halo mass, which aligns with our findings. In contrast, \cite{Anderson_2015_MNRAS.449.3806A} focused on locally brightest galaxies rather than galaxy groups or clusters. Their selection included galaxies with stellar masses ranging from $10^{10}$ to $10^{12}M_{\odot}$, corresponding to halos with masses comparable to those in our sample. The luminosity estimates in their work, ranging from $10^{40}$ to $10^{42}$ erg/s, are consistent with ours. Only a few subsamples with the largest stellar mass ranges in their study exceed this range, reaching up to $10^{43}$ erg/s.

\subsubsection{Comparison with Individual Source Studies}

In addition to the stacking-based studies discussed above, we also compare our results with studies focusing on individual sources. We collected results from literatures \citep{Mulchaey_96_APJ, Finoguenov_07_APJS, Sun_09_APJ, Poon_23_MNRAS} and presented them in Fig. \ref{fig:comparison} along with our findings (shown in red points). To ensure consistency, we recalculated the halo mass and corresponding measurements from within the range of $r_{180}$ to $r_{500}$. We assumed a Navarro-Frenk-White (NFW) density profile \citep{NFW_97_APJ}, with concentration parameters given by \cite{Maccio_07_MNRAS}. We obtained $M_\mathrm{halo}/M_\mathrm{500}$ = 1.38 and $r_\mathrm{180}/r_\mathrm{500}$ = 1.56 when the concentration index is $c_\mathrm{180}$ = 6. We tried different concentration index values between 4--12 and found no significant differences in results. With these conversions, we recalculated the luminosity and gas mass for the poor group samples within $r_\mathrm{500}$. 

These previous results encompassed various systems, including poor galaxy groups identified in \cite{Mulchaey_96_APJ} (green points), rich galaxy groups and clusters (blue points). To better compare with our results, we included both the $L_\mathrm{X}$-$M_\mathrm{500}$ relation based on galaxy clusters from \cite{Schellenberger_2017_MNRAS} (dashed line) and the $L_\mathrm{X}$-$M_\mathrm{500}$ relation based on galaxy groups from \cite{Lovisari_2015_A&A...573A.118L} (solid line) in the left panel of Fig. \ref{fig:comparison}. Our results show some offsets from both relations. The large error bars and possible contamination in the luminosity measurements (discussed later) make it challenging to establish a definitive $L_\mathrm{X}$-$M_\mathrm{500}$ scaling relation for poor groups at this stage. Further investigations, with more precise measurements and better constraints, are required to clarify this aspect.

A pivotal aspect of this study is the estimation of baryon content in poor groups. We estimated the typical stellar mass for each subsample using the $M_*$--$M_{500}$ relation from the IllustrisTNG simulation \citep{Pillepich_2018_MNRAS} and summed the gas and stellar masses to calculate the total baryon mass. We found that the average baryon fraction within $r_{500}$ is approximately 8\%, which is significantly lower than the 16\% expected from the cosmic mean \citep{Planck18_20_A&A}. This strongly implies the persistence of the ``missing baryons'' problem in poor groups. As discussed in \cite{Mulchaey_00_ARAA}, due to their limited gravitational potential, it is possible that a substantial amount of gas within $r_{500}$ has been expelled to larger radii. Additionally, a significant portion of the gas may exist in a lower-temperature phase, emitting at lower-energy X-ray or UV wavelengths, which would require observations in these bands for detection. Further studies are needed to investigate these possibilities.

The average baryon fractions within the four halo mass ranges ($10^{11.5}$--$10^{12.0}$, $10^{12.0}$--$10^{12.5}$, $10^{12.5}$--$10^{13.0}$, and $10^{13.0}$--$10^{13.5} h^{-1}M_\mathrm{\odot}$) are 12\%, 10\%, 7\%, and 6\%, respectively, which seems to indicate that the ``missing baryons'' problem is less severe at the lower mass end. This may be due to two reasons. On the one hand, data at the low-mass end has a lower S/N and a higher proportion of contamination, which may lead to an overestimation of the gas mass. On the other hand, it is possible that in poor groups at the low-mass end, the supermassive black holes are relatively smaller in size, the AGN feedback is weaker, and more gas is retained within the poor groups. The declining trend in baryon fraction from $10^{11.5}$ to $10^{13.5}$ $M_{\mathrm{\odot}}$ is also corroborated by results from the IllustrisTNG simulation (Hong-Chuan Ma, private communication). Further research is needed to confirm these possibilities.

\subsection{Impact of Temperature and Metallicity Uncertainties on Baryon Fractions}
Based on the $M-T$ and $M-Z$ relations \citep{Sun_09_APJ,Truong_2019_MNRAS}, we can derive four temperature and metallicity values corresponding to the four halo mass bins in this work. The temperatures are 0.08, 0.12, 0.25, and 0.48 keV respectively, while the metallicities are 0.65, 0.60, 0.54, and 0.47 $Z_{\odot}$. To address the potential impact on baryon mass due to uncertainties in temperature and metallicity, we adopted two schemes for combining these parameters to calculate the baryon mass.

The first scheme involves fixing the temperature for each of the four halo mass bins to one of the aforementioned values while allowing the metallicity to vary across the four different values, or vice versa. This approach results in a total of 8 possible combinations. The second scheme entails setting both the temperature and metallicity values for the four halo mass bins to the same fixed value, resulting in 16 possible combinations.

Under each combination, we can calculate an average baryon fraction for a set of 12 subsamples. We ultimately find that across all these combinations, the highest average baryon fraction obtained is approximately 10\%, and the lowest is around 6\%. These results suggest that our main conclusions are not significantly affected by the uncertainties in temperature and metallicity.

\begin{figure}[tbp]
    \centering
    \includegraphics[width=0.48\textwidth, keepaspectratio]{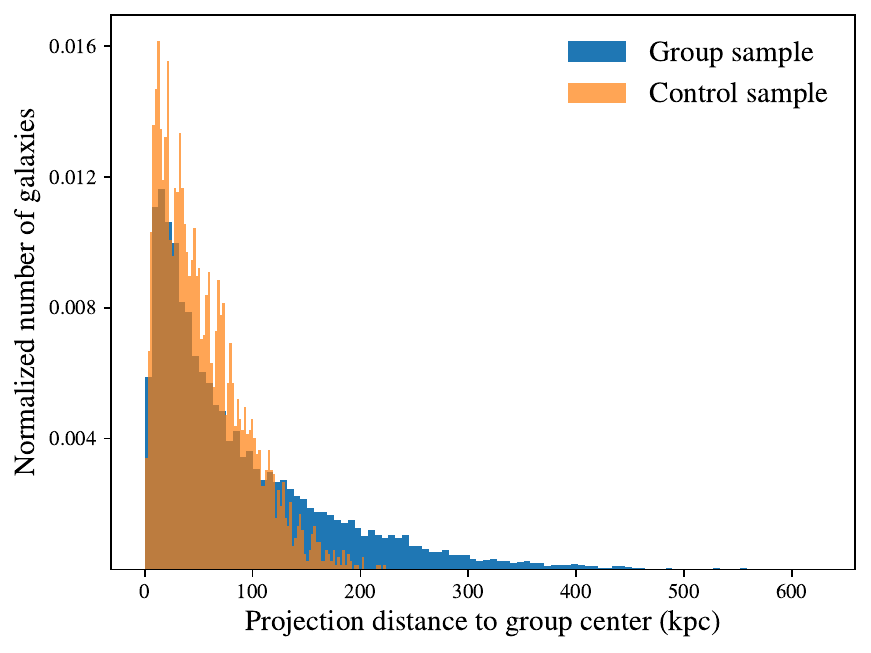}
    \caption{The distance distribution of member galaxies to the group centers in the actual group samples and control samples. The group centers were defined based on DESI LS data as the luminosity-weighted centers of the member galaxies, for both real and mock groups. Both samples were normalized.}
    \label{fig:control_center}
\end{figure}

\subsection{Contamination from Unresolved Point Sources}
Unresolved point sources, including X-ray binaries (XRBs) and low-luminosity active galactic nuclei (AGNs), remain in the stacked emission as contamination. We estimated the luminosity of these point sources using empirical formulas, along with photometric data of member galaxies from DESI LS.

We applied empirical models to estimate the contaminations from unresolved point sources \citep[e.g.,][]{Aird_2017_MNRAS, Comparat_22_A&A}, which require information on the stellar mass, redshift, and star formation rate (SFR) of the member galaxies. We have direct stellar mass information for only about 20\% of the galaxies from DESI LS public data. For the remaining 80\%, we estimated stellar masses using common stellar-to-halo mass ratios provided by \cite{Yang_2007_ApJ} for specific halo mass ranges. Lacking direct SFR data, we adopted a uniform SFR assumption of 5 $M_{\odot}$ yr$^{-1}$, reflecting the typical range of 1--10 $M_{\odot}$ yr$^{-1}$ for star-forming galaxies \citep[e.g.,][]{Salim_2007_ApJS..173..267S}. To ensure robustness, we tested an alternative value of 10 $M_{\odot}$ yr$^{-1}$ and found no significant impact on the final contamination level, except for the lowest mass subsample, where it increases substantially. For other subsamples, the contamination level increases only slightly by a few percent.

For XRBs luminosity in the 2--10 keV band, we followed the model by \cite{Aird_2017_MNRAS}, which correlates XRB luminosity with SFR and stellar mass. We converted this luminosity to the 0.5--2.0 keV band using a power law model with a photon index of 1.8, and considered the median XRBs luminosity of all groups within each subsample as the representative XRBs contamination.

There was an investigation into the point-source emission emerging from GAMA galaxy stacks to determine the faint end of the AGN X-ray luminosity function \citep[referred to][]{Comparat_22_A&A}. They quantitatively assessed both the average X-ray luminosity ($L_X$) and the prevalence of X-ray AGN across different stellar mass ranges. Specifically, for galaxies with a stellar mass (expressed as log($M_{*}/[M_{\odot}]$)) of 9.75, 10.75, and 11.75, the occupation fraction of galaxies hosting an X-ray AGN is found to be 0.1\%, 1\%, and 10\%, respectively. Corresponding to these occupation fractions, the average X-ray luminosities ($L_X$) were determined to be log($L_X/[\mathrm{erg\, s^{-1}]}$) of approximately 40, 41, and 42. By multiplying these average luminosities by their corresponding occupation fractions, we derived the expected AGN luminosity for member galaxies based on their stellar mass. Summing the AGN contributions from all member galaxies within a group provides an estimate of that group's total AGN contamination. We then determined the median AGN contamination across all groups within each subsample to represent its characteristic AGN contamination level.

We defined the contamination fraction as the ratio of total XRBs and AGNs luminosities to the luminosity of the total stacked emission. Our findings indicated an average contamination fraction of about 20\% across detected subsamples. The subsample with a redshift range of 0.1--0.2 and a halo mass range of $10^{11.5}$--$10^{12.0} h^{-1}M_\mathrm{\odot}$ shows a contamination fraction nearing 50\%, likely due to its low S/N or comparatively lesser hot gas content in lower mass groups. Conversely, higher halo mass ranges ($10^{12.0}$--$10^{12.5}$, $10^{12.5}$--$10^{13.0}$, and $10^{13.0}$--$10^{13.5} h^{-1}M_\mathrm{\odot}$) exhibit lower contamination fractions of 30\%, 12\%, and 8\%, respectively, suggesting a trend of decreasing contamination with increasing halo mass.

\subsection{Potential Contamination from the CGM}
The hot CGM of member galaxies is another possible source of contamination. To validate the robustness of the detection of hot IGrM in poor groups, We constructed a control sample to estimate the contamination from the hot CGM of individual galaxies. Specifically, we first selected isolated galaxies from \cite{Yang_2021_APJ}'s catalog. These isolated galaxies were then used to construct mock galaxy groups, with each mock group consisting of 2 to 4 isolated galaxies. The total halo mass and redshifts of these mock galaxy groups were selected to match the same range as those of the real group samples. Additionally, the galaxies within each mock group were required to be located within a projected radial distance of fewer than 300 kiloparsecs from one another. We matched the number of mock groups in each subsample to that of the corresponding real group subsample to ensure consistent data depth. Since these mock groups do not represent physical systems, they do not contain the hot IGrM. Any emission detected after performing stacking analysis on these mock groups is likely to originate from individual galaxies. \cite{Yang_2021_APJ} defined the center of a real group as the luminosity-weighted center of its member galaxies based on DESI LS data, and we adopted the same definition for the centers of mock groups. Upon stacking these 16 control subsamples individually, we did not find any signal above the 2$\sigma$ level. We display two stacked images of control sample as examples in Fig. \ref{fig:control_image}. To enhance the reliability of the control sample, we adjusted the distance distribution of member galaxies to their mock group centers to align with that of the actual poor group sample, as illustrated in Fig. \ref{fig:control_center}.

\section{Summary} \label{sec:summary}
In this work, we conducted an investigation into the properties of the extended hot IGrM within poor galaxy groups using eROSITA eFEDS data. We compiled a sample of 25,524 poor group samples and categorized them into 16 subsamples based on their redshift and halo mass. Subsequently, we performed stacking analyses to measure the mean luminosity, gas mass, and gas fraction of these stacked groups, and compared our findings with previous studies. We revealed the ubiquitous presence of the hot IGrM in an optically selected, poor galaxy group sample and assessed the average luminosity and baryon mass. The baryon fraction of poor groups is significantly lower than the predicted values by the $\Lambda$CDM model. This indicates that there is still an ``missing baryons'' problem in poor groups. In the future, advanced X-ray instruments such as HUBS \citep{HUBS_2023_SCPMA} and Athena \citep{Athena_2013_arXiv1306.2307N} will provide excellent opportunities to observe the hot IGrM of individual poor groups in the nearby universe within reasonable exposure times. In the Local Group, the hot gas likely interacts with or mixes with that of the Milky Way, complicating detection via X-ray emission. Nevertheless, X-ray absorption techniques may offer a promising approach to resolve the hot IGrM of the Local Group.

\section*{Acknowledgments}
We thank the anonymous referee for valuable comments. This work is supported by the National Natural Science Foundation of China (NSFC) under Nos. 11890692, 12133008, 12221003, 12373007, 12422302. We acknowledge the science research grants from the China Manned
Space Project with No. CMS-CSST-2021-A04. FN acknowledges support from the INAF-Fundamental-Astrophysics grant No. 1.05.23.01.06 and the Italian Ministry of University and Research PRIN-2022 grant No. 2.06.01.20.

This work is based on data from eROSITA, the soft X-ray instrument aboard SRG, a joint Russian-German science mission supported by the Russian Space Agency (Roskosmos), in the interests of the Russian Academy of Sciences represented by its Space Research Institute (IKI), and the Deutsches Zentrum f{\"u}r Luft- und Raumfahrt (DLR). The SRG spacecraft was built by Lavochkin Association (NPOL) and its subcontractors, and is operated by NPOL with support from the Max Planck Institute for Extraterrestrial Physics (MPE). The development and construction of the eROSITA X-ray instrument was led by MPE, with contributions from the Dr. Karl Remeis Observatory Bamberg \& ECAP (FAU Erlangen-Nuernberg), the University of Hamburg Observatory, the Leibniz Institute for Astrophysics Potsdam (AIP), and the Institute for Astronomy and Astrophysics of the University of T{\"u}bingen, with the support of DLR and the Max Planck Society. The Argelander Institute for Astronomy of the University of Bonn and the Ludwig Maximilians Universit{\"a}t Munich also participated in the science preparation for eROSITA. The eROSITA data shown here were processed using the eSASS software system developed by the German eROSITA consortium.


\bibliographystyle{aasjournal}
\bibliography{ref}

\begin{thebibliography}{}
\expandafter\ifx\csname natexlab\endcsname\relax\def\natexlab#1{#1}\fi
\providecommand{\url}[1]{\href{#1}{#1}}
\providecommand{\dodoi}[1]{doi:~\href{http://doi.org/#1}{\nolinkurl{#1}}}
\providecommand{\doeprint}[1]{\href{http://ascl.net/#1}{\nolinkurl{http://ascl.net/#1}}}
\providecommand{\doarXiv}[1]{\href{https://arxiv.org/abs/#1}{\nolinkurl{https://arxiv.org/abs/#1}}}

\bibitem[{{Aird} {et~al.}(2017){Aird}, {Coil}, \&
  {Georgakakis}}]{Aird_2017_MNRAS}
{Aird}, J., {Coil}, A.~L., \& {Georgakakis}, A. 2017, \mnras, 465, 3390,
  \dodoi{10.1093/mnras/stw2932}

\bibitem[{{Anderson} {et~al.}(2015){Anderson}, {Gaspari}, {White}, {Wang}, \&
  {Dai}}]{Anderson_2015_MNRAS.449.3806A}
{Anderson}, M.~E., {Gaspari}, M., {White}, S. D.~M., {Wang}, W., \& {Dai}, X.
  2015, \mnras, 449, 3806, \dodoi{10.1093/mnras/stv437}

\bibitem[{{Andrae}(2010)}]{Andrae_2010_arXiv1009.2755A}
{Andrae}, R. 2010, arXiv e-prints, arXiv:1009.2755,
  \dodoi{10.48550/arXiv.1009.2755}

\bibitem[{{Bregman} {et~al.}(2023){Bregman}, {Cen}, {Chen}, {Cui}, {Fang},
  {Guo}, {Hodges-Kluck}, {Huang}, {Ho}, {Ji}, {Ji}, {Kang}, {Lai}, {Li}, {Li},
  {Li}, {Li}, {Li}, {Li}, {Liang}, {Liu}, {Liu}, {Lu}, {Mao}, {Ponti}, {Qu},
  {Shan}, {Shao}, {Shi}, {Shu}, {Sun}, {Sun}, {Tong}, {Wang}, {Wang}, {Wang},
  {Wang}, {Wang}, {Wang}, {Wang}, {Xu}, {Xu}, {Xu}, {Xu}, {Xu}, {Xue}, {Yang},
  {Yuan}, {Zhang}, {Zhang}, {Zhang}, {Zhao}, {Zhou}, \&
  {Zhou}}]{HUBS_2023_SCPMA}
{Bregman}, J., {Cen}, R., {Chen}, Y., {et~al.} 2023, Science China Physics,
  Mechanics, and Astronomy, 66, 299513, \dodoi{10.1007/s11433-023-2149-y}

\bibitem[{{Brunner} {et~al.}(2022){Brunner}, {Liu}, {Lamer}, {Georgakakis},
  {Merloni}, {Brusa}, {Bulbul}, {Dennerl}, {Friedrich}, {Liu}, {Maitra},
  {Nandra}, {Ramos-Ceja}, {Sanders}, {Stewart}, {Boller}, {Buchner}, {Clerc},
  {Comparat}, {Dwelly}, {Eckert}, {Finoguenov}, {Freyberg}, {Ghirardini},
  {Gueguen}, {Haberl}, {Kreykenbohm}, {Krumpe}, {Osterhage}, {Pacaud},
  {Predehl}, {Reiprich}, {Robrade}, {Salvato}, {Santangelo}, {Schrabback},
  {Schwope}, \& {Wilms}}]{Brunner_2022_A&A}
{Brunner}, H., {Liu}, T., {Lamer}, G., {et~al.} 2022, \aap, 661, A1,
  \dodoi{10.1051/0004-6361/202141266}

\bibitem[{{Cavaliere} \& {Fusco-Femiano}(1976)}]{Cavaliere_76_A&A}
{Cavaliere}, A., \& {Fusco-Femiano}, R. 1976, \aap, 49, 137

\bibitem[{{Cen} \& {Ostriker}(2006)}]{Cen_2006_ApJ...650..560C}
{Cen}, R., \& {Ostriker}, J.~P. 2006, \apj, 650, 560, \dodoi{10.1086/506505}

\bibitem[{{Chadayammuri} {et~al.}(2022){Chadayammuri}, {Bogd{\'a}n},
  {Oppenheimer}, {Kraft}, {Forman}, \& {Jones}}]{Chadayammuri_22_APJL}
{Chadayammuri}, U., {Bogd{\'a}n}, {\'A}., {Oppenheimer}, B.~D., {et~al.} 2022,
  \apjl, 936, L15, \dodoi{10.3847/2041-8213/ac8936}

\bibitem[{{Cole} \& {Kaiser}(1988)}]{Cole_1988_MNRAS}
{Cole}, S., \& {Kaiser}, N. 1988, \mnras, 233, 637,
  \dodoi{10.1093/mnras/233.3.637}

\bibitem[{{Comparat} {et~al.}(2022){Comparat}, {Truong}, {Merloni},
  {Pillepich}, {Ponti}, {Driver}, {Bellstedt}, {Liske}, {Aird}, {Br{\"u}ggen},
  {Bulbul}, {Davies}, {Villalba}, {Georgakakis}, {Haberl}, {Liu}, {Maitra},
  {Nandra}, {Popesso}, {Predehl}, {Robotham}, {Salvato}, {Thorne}, \&
  {Zhang}}]{Comparat_22_A&A}
{Comparat}, J., {Truong}, N., {Merloni}, A., {et~al.} 2022, \aap, 666, A156,
  \dodoi{10.1051/0004-6361/202243101}

\bibitem[{{Dai} {et~al.}(2007){Dai}, {Kochanek}, \& {Morgan}}]{Dai_07_APJ}
{Dai}, X., {Kochanek}, C.~S., \& {Morgan}, N.~D. 2007, \apj, 658, 917,
  \dodoi{10.1086/509651}

\bibitem[{{Dav{\'e}} {et~al.}(2001){Dav{\'e}}, {Cen}, {Ostriker}, {Bryan},
  {Hernquist}, {Katz}, {Weinberg}, {Norman}, \&
  {O'Shea}}]{Cen_2001_ApJ...552..473D}
{Dav{\'e}}, R., {Cen}, R., {Ostriker}, J.~P., {et~al.} 2001, \apj, 552, 473,
  \dodoi{10.1086/320548}

\bibitem[{{Dey} {et~al.}(2019){Dey}, {Schlegel}, {Lang}, {Blum}, {Burleigh},
  {Fan}, {Findlay}, {Finkbeiner}, {Herrera}, {Juneau}, {Landriau}, {Levi},
  {McGreer}, {Meisner}, {Myers}, {Moustakas}, {Nugent}, {Patej}, {Schlafly},
  {Walker}, {Valdes}, {Weaver}, {Y{\`e}che}, {Zou}, {Zhou}, {Abareshi},
  {Abbott}, {Abolfathi}, {Aguilera}, {Alam}, {Allen}, {Alvarez}, {Annis},
  {Ansarinejad}, {Aubert}, {Beechert}, {Bell}, {BenZvi}, {Beutler}, {Bielby},
  {Bolton}, {Brice{\~n}o}, {Buckley-Geer}, {Butler}, {Calamida}, {Carlberg},
  {Carter}, {Casas}, {Castander}, {Choi}, {Comparat}, {Cukanovaite}, {Delubac},
  {DeVries}, {Dey}, {Dhungana}, {Dickinson}, {Ding}, {Donaldson}, {Duan},
  {Duckworth}, {Eftekharzadeh}, {Eisenstein}, {Etourneau}, {Fagrelius},
  {Farihi}, {Fitzpatrick}, {Font-Ribera}, {Fulmer}, {G{\"a}nsicke},
  {Gaztanaga}, {George}, {Gerdes}, {Gontcho}, {Gorgoni}, {Green}, {Guy},
  {Harmer}, {Hernandez}, {Honscheid}, {Huang}, {James}, {Jannuzi}, {Jiang},
  {Joyce}, {Karcher}, {Karkar}, {Kehoe}, {Kneib}, {Kueter-Young}, {Lan},
  {Lauer}, {Le Guillou}, {Le Van Suu}, {Lee}, {Lesser}, {Perreault Levasseur},
  {Li}, {Mann}, {Marshall}, {Mart{\'\i}nez-V{\'a}zquez}, {Martini}, {du Mas des
  Bourboux}, {McManus}, {Meier}, {M{\'e}nard}, {Metcalfe},
  {Mu{\~n}oz-Guti{\'e}rrez}, {Najita}, {Napier}, {Narayan}, {Newman}, {Nie},
  {Nord}, {Norman}, {Olsen}, {Paat}, {Palanque-Delabrouille}, {Peng},
  {Poppett}, {Poremba}, {Prakash}, {Rabinowitz}, {Raichoor}, {Rezaie},
  {Robertson}, {Roe}, {Ross}, {Ross}, {Rudnick}, {Safonova}, {Saha},
  {S{\'a}nchez}, {Savary}, {Schweiker}, {Scott}, {Seo}, {Shan}, {Silva},
  {Slepian}, {Soto}, {Sprayberry}, {Staten}, {Stillman}, {Stupak}, {Summers},
  {Sien Tie}, {Tirado}, {Vargas-Maga{\~n}a}, {Vivas}, {Wechsler}, {Williams},
  {Yang}, {Yang}, {Yapici}, {Zaritsky}, {Zenteno}, {Zhang}, {Zhang}, {Zhou}, \&
  {Zhou}}]{DESI_2019_AJ}
{Dey}, A., {Schlegel}, D.~J., {Lang}, D., {et~al.} 2019, \aj, 157, 168,
  \dodoi{10.3847/1538-3881/ab089d}

\bibitem[{{Eke} {et~al.}(2004){Eke}, {Baugh}, {Cole}, {Frenk}, {Norberg},
  {Peacock}, {Baldry}, {Bland-Hawthorn}, {Bridges}, {Cannon}, {Colless},
  {Collins}, {Couch}, {Dalton}, {de Propris}, {Driver}, {Efstathiou}, {Ellis},
  {Glazebrook}, {Jackson}, {Lahav}, {Lewis}, {Lumsden}, {Maddox}, {Madgwick},
  {Peterson}, {Sutherland}, \& {Taylor}}]{Eke_04_MNRAS}
{Eke}, V.~R., {Baugh}, C.~M., {Cole}, S., {et~al.} 2004, \mnras, 348, 866,
  \dodoi{10.1111/j.1365-2966.2004.07408.x}

\bibitem[{{Finoguenov} {et~al.}(2007){Finoguenov}, {Guzzo}, {Hasinger},
  {Scoville}, {Aussel}, {B{\"o}hringer}, {Brusa}, {Capak}, {Cappelluti},
  {Comastri}, {Giodini}, {Griffiths}, {Impey}, {Koekemoer}, {Kneib},
  {Leauthaud}, {Le F{\`e}vre}, {Lilly}, {Mainieri}, {Massey}, {McCracken},
  {Mobasher}, {Murayama}, {Peacock}, {Sakelliou}, {Schinnerer}, {Silverman},
  {Smol{\v{c}}i{\'c}}, {Taniguchi}, {Tasca}, {Taylor}, {Trump}, \&
  {Zamorani}}]{Finoguenov_07_APJS}
{Finoguenov}, A., {Guzzo}, L., {Hasinger}, G., {et~al.} 2007, \apjs, 172, 182,
  \dodoi{10.1086/516577}

\bibitem[{{Fukugita} {et~al.}(1998){Fukugita}, {Hogan}, \&
  {Peebles}}]{Fukugita_1998_ApJ}
{Fukugita}, M., {Hogan}, C.~J., \& {Peebles}, P.~J.~E. 1998, \apj, 503, 518,
  \dodoi{10.1086/306025}

\bibitem[{{Ge} {et~al.}(2016){Ge}, {Wang}, {Tripp}, {Li}, {Gu}, \&
  {Ji}}]{Ge_16_MNRAS}
{Ge}, C., {Wang}, Q.~D., {Tripp}, T.~M., {et~al.} 2016, \mnras, 459, 366,
  \dodoi{10.1093/mnras/stw599}

\bibitem[{{Helsdon} \& {Ponman}(2000)}]{Helsdon_2000_MNRAS}
{Helsdon}, S.~F., \& {Ponman}, T.~J. 2000, \mnras, 315, 356,
  \dodoi{10.1046/j.1365-8711.2000.03396.x}

\bibitem[{{Helsdon} {et~al.}(2005){Helsdon}, {Ponman}, \&
  {Mulchaey}}]{Helsdon_2005_ApJ}
{Helsdon}, S.~F., {Ponman}, T.~J., \& {Mulchaey}, J.~S. 2005, \apj, 618, 679,
  \dodoi{10.1086/426009}

\bibitem[{{Liu} {et~al.}(2022){Liu}, {Bulbul}, {Ghirardini}, {Liu}, {Klein},
  {Clerc}, {{\"O}zsoy}, {Ramos-Ceja}, {Pacaud}, {Comparat}, {Okabe}, {Bahar},
  {Biffi}, {Brunner}, {Br{\"u}ggen}, {Buchner}, {Ider Chitham}, {Chiu},
  {Dolag}, {Gatuzz}, {Gonzalez}, {Hoang}, {Lamer}, {Merloni}, {Nandra},
  {Oguri}, {Ota}, {Predehl}, {Reiprich}, {Salvato}, {Schrabback}, {Sanders},
  {Seppi}, \& {Thibaud}}]{Liu_Ang_22_A&A}
{Liu}, A., {Bulbul}, E., {Ghirardini}, V., {et~al.} 2022, \aap, 661, A2,
  \dodoi{10.1051/0004-6361/202141120}

\bibitem[{{Lovisari} {et~al.}(2015){Lovisari}, {Reiprich}, \&
  {Schellenberger}}]{Lovisari_2015_A&A...573A.118L}
{Lovisari}, L., {Reiprich}, T.~H., \& {Schellenberger}, G. 2015, \aap, 573,
  A118, \dodoi{10.1051/0004-6361/201423954}

\bibitem[{{Macci{\`o}} {et~al.}(2007){Macci{\`o}}, {Dutton}, {van den Bosch},
  {Moore}, {Potter}, \& {Stadel}}]{Maccio_07_MNRAS}
{Macci{\`o}}, A.~V., {Dutton}, A.~A., {van den Bosch}, F.~C., {et~al.} 2007,
  \mnras, 378, 55, \dodoi{10.1111/j.1365-2966.2007.11720.x}

\bibitem[{{McGaugh} {et~al.}(2010){McGaugh}, {Schombert}, {de Blok}, \&
  {Zagursky}}]{McGaugh_2010_ApJ}
{McGaugh}, S.~S., {Schombert}, J.~M., {de Blok}, W.~J.~G., \& {Zagursky}, M.~J.
  2010, \apjl, 708, L14, \dodoi{10.1088/2041-8205/708/1/L14}

\bibitem[{{McIntosh}(2016)}]{McIntosh_2016_arXiv160600497M}
{McIntosh}, A. 2016, arXiv e-prints, arXiv:1606.00497,
  \dodoi{10.48550/arXiv.1606.00497}

\bibitem[{{Merloni} {et~al.}(2020){Merloni}, {Nandra}, \&
  {Predehl}}]{Merloni_20_NA}
{Merloni}, A., {Nandra}, K., \& {Predehl}, P. 2020, Nature Astronomy, 4, 634,
  \dodoi{10.1038/s41550-020-1133-0}

\bibitem[{{Merloni} {et~al.}(2024){Merloni}, {Lamer}, {Liu}, {Ramos-Ceja},
  {Brunner}, {Bulbul}, {Dennerl}, {Doroshenko}, {Freyberg}, {Friedrich},
  {Gatuzz}, {Georgakakis}, {Haberl}, {Igo}, {Kreykenbohm}, {Liu}, {Maitra},
  {Malyali}, {Mayer}, {Nandra}, {Predehl}, {Robrade}, {Salvato}, {Sanders},
  {Stewart}, {Tub{\'\i}n-Arenas}, {Weber}, {Wilms}, {Arcodia}, {Artis},
  {Aschersleben}, {Avakyan}, {Aydar}, {Bahar}, {Balzer}, {Becker}, {Berger},
  {Boller}, {Bornemann}, {Br{\"u}ggen}, {Brusa}, {Buchner}, {Burwitz},
  {Camilloni}, {Clerc}, {Comparat}, {Coutinho}, {Czesla}, {Dannhauer},
  {Dauner}, {Dauser}, {Dietl}, {Dolag}, {Dwelly}, {Egg}, {Ehl}, {Freund},
  {Friedrich}, {Gaida}, {Garrel}, {Ghirardini}, {Gokus}, {Gr{\"u}nwald},
  {Grandis}, {Grotova}, {Gruen}, {Gueguen}, {H{\"a}mmerich}, {Hamaus},
  {Hasinger}, {Haubner}, {Homan}, {Ider Chitham}, {Joseph}, {Joyce},
  {K{\"o}nig}, {Kaltenbrunner}, {Khokhriakova}, {Kink}, {Kirsch}, {Kluge},
  {Knies}, {Krippendorf}, {Krumpe}, {Kurpas}, {Li}, {Liu}, {Locatelli},
  {Lorenz}, {M{\"u}ller}, {Magaudda}, {Mannes}, {McCall}, {Meidinger},
  {Michailidis}, {Migkas}, {Mu{\~n}oz-Giraldo}, {Musiimenta}, {Nguyen-Dang},
  {Ni}, {Olechowska}, {Ota}, {Pacaud}, {Pasini}, {Perinati}, {Pires},
  {Pommranz}, {Ponti}, {Poppenhaeger}, {P{\"u}hlhofer}, {Rau}, {Reh},
  {Reiprich}, {Roster}, {Saeedi}, {Santangelo}, {Sasaki}, {Schmitt},
  {Schneider}, {Schrabback}, {Schuster}, {Schwope}, {Seppi}, {Serim},
  {Shreeram}, {Sokolova-Lapa}, {Starck}, {Stelzer}, {Stierhof}, {Suleimanov},
  {Tenzer}, {Traulsen}, {Tr{\"u}mper}, {Tsuge}, {Urrutia}, {Veronica},
  {Waddell}, {Willer}, {Wolf}, {Yeung}, {Zainab}, {Zangrandi}, {Zhang},
  {Zhang}, \& {Zheng}}]{Merloni_2024_A&A}
{Merloni}, A., {Lamer}, G., {Liu}, T., {et~al.} 2024, \aap, 682, A34,
  \dodoi{10.1051/0004-6361/202347165}

\bibitem[{{Mohr} {et~al.}(1999){Mohr}, {Mathiesen}, \&
  {Evrard}}]{Mohr_1999_ApJ}
{Mohr}, J.~J., {Mathiesen}, B., \& {Evrard}, A.~E. 1999, \apj, 517, 627,
  \dodoi{10.1086/307227}

\bibitem[{{Mulchaey}(2000)}]{Mulchaey_00_ARAA}
{Mulchaey}, J.~S. 2000, \araa, 38, 289, \dodoi{10.1146/annurev.astro.38.1.289}

\bibitem[{{Mulchaey} {et~al.}(1996){Mulchaey}, {Davis}, {Mushotzky}, \&
  {Burstein}}]{Mulchaey_96_APJ}
{Mulchaey}, J.~S., {Davis}, D.~S., {Mushotzky}, R.~F., \& {Burstein}, D. 1996,
  \apj, 456, 80, \dodoi{10.1086/176629}

\bibitem[{{Mulchaey} \& {Zabludoff}(1998)}]{Mulchaey_98_APJ}
{Mulchaey}, J.~S., \& {Zabludoff}, A.~I. 1998, \apj, 496, 73,
  \dodoi{10.1086/305356}

\bibitem[{{Nandra} {et~al.}(2013){Nandra}, {Barret}, {Barcons}, {Fabian}, {den
  Herder}, {Piro}, {Watson}, {Adami}, {Aird}, {Afonso}, {Alexander},
  {Argiroffi}, {Amati}, {Arnaud}, {Atteia}, {Audard}, {Badenes}, {Ballet},
  {Ballo}, {Bamba}, {Bhardwaj}, {Stefano Battistelli}, {Becker}, {De Becker},
  {Behar}, {Bianchi}, {Biffi}, {B{\^\i}rzan}, {Bocchino}, {Bogdanov}, {Boirin},
  {Boller}, {Borgani}, {Borm}, {Bouch{\'e}}, {Bourdin}, {Bower}, {Braito},
  {Branchini}, {Branduardi-Raymont}, {Bregman}, {Brenneman}, {Brightman},
  {Br{\"u}ggen}, {Buchner}, {Bulbul}, {Brusa}, {Bursa}, {Caccianiga},
  {Cackett}, {Campana}, {Cappelluti}, {Cappi}, {Carrera}, {Ceballos},
  {Christensen}, {Chu}, {Churazov}, {Clerc}, {Corbel}, {Corral}, {Comastri},
  {Costantini}, {Croston}, {Dadina}, {D'Ai}, {Decourchelle}, {Della Ceca},
  {Dennerl}, {Dolag}, {Done}, {Dovciak}, {Drake}, {Eckert}, {Edge}, {Ettori},
  {Ezoe}, {Feigelson}, {Fender}, {Feruglio}, {Finoguenov}, {Fiore}, {Galeazzi},
  {Gallagher}, {Gandhi}, {Gaspari}, {Gastaldello}, {Georgakakis},
  {Georgantopoulos}, {Gilfanov}, {Gitti}, {Gladstone}, {Goosmann}, {Gosset},
  {Grosso}, {Guedel}, {Guerrero}, {Haberl}, {Hardcastle}, {Heinz}, {Alonso
  Herrero}, {Herv{\'e}}, {Holmstrom}, {Iwasawa}, {Jonker}, {Kaastra}, {Kara},
  {Karas}, {Kastner}, {King}, {Kosenko}, {Koutroumpa}, {Kraft}, {Kreykenbohm},
  {Lallement}, {Lanzuisi}, {Lee}, {Lemoine-Goumard}, {Lobban}, {Lodato},
  {Lovisari}, {Lotti}, {McCharthy}, {McNamara}, {Maggio}, {Maiolino}, {De
  Marco}, {de Martino}, {Mateos}, {Matt}, {Maughan}, {Mazzotta}, {Mendez},
  {Merloni}, {Micela}, {Miceli}, {Mignani}, {Miller}, {Miniutti}, {Molendi},
  {Montez}, {Moretti}, {Motch}, {Naz{\'e}}, {Nevalainen}, {Nicastro}, {Nulsen},
  {Ohashi}, {O'Brien}, {Osborne}, {Oskinova}, {Pacaud}, {Paerels}, {Page},
  {Papadakis}, {Pareschi}, {Petre}, {Petrucci}, {Piconcelli}, {Pillitteri},
  {Pinto}, {de Plaa}, {Pointecouteau}, {Ponman}, {Ponti}, {Porquet}, {Pounds},
  {Pratt}, {Predehl}, {Proga}, {Psaltis}, {Rafferty}, {Ramos-Ceja}, {Ranalli},
  {Rasia}, {Rau}, {Rauw}, {Rea}, {Read}, {Reeves}, {Reiprich}, {Renaud},
  {Reynolds}, {Risaliti}, {Rodriguez}, {Rodriguez Hidalgo}, {Roncarelli},
  {Rosario}, {Rossetti}, {Rozanska}, {Rovilos}, {Salvaterra}, {Salvato}, {Di
  Salvo}, {Sanders}, {Sanz-Forcada}, {Schawinski}, {Schaye}, {Schwope},
  {Sciortino}, {Severgnini}, {Shankar}, {Sijacki}, {Sim}, {Schmid}, {Smith},
  {Steiner}, {Stelzer}, {Stewart}, {Strohmayer}, {Str{\"u}der}, {Sun}, {Takei},
  {Tatischeff}, {Tiengo}, {Tombesi}, {Trinchieri}, {Tsuru}, {Ud-Doula},
  {Ursino}, {Valencic}, {Vanzella}, {Vaughan}, {Vignali}, {Vink}, {Vito},
  {Volonteri}, {Wang}, {Webb}, {Willingale}, {Wilms}, {Wise}, {Worrall},
  {Young}, {Zampieri}, {In't Zand}, {Zane}, {Zezas}, {Zhang}, \&
  {Zhuravleva}}]{Athena_2013_arXiv1306.2307N}
{Nandra}, K., {Barret}, D., {Barcons}, X., {et~al.} 2013, arXiv e-prints,
  arXiv:1306.2307, \dodoi{10.48550/arXiv.1306.2307}

\bibitem[{{Navarro} {et~al.}(1997){Navarro}, {Frenk}, \& {White}}]{NFW_97_APJ}
{Navarro}, J.~F., {Frenk}, C.~S., \& {White}, S. D.~M. 1997, \apj, 490, 493,
  \dodoi{10.1086/304888}

\bibitem[{{Pillepich} {et~al.}(2018){Pillepich}, {Nelson}, {Hernquist},
  {Springel}, {Pakmor}, {Torrey}, {Weinberger}, {Genel}, {Naiman}, {Marinacci},
  \& {Vogelsberger}}]{Pillepich_2018_MNRAS}
{Pillepich}, A., {Nelson}, D., {Hernquist}, L., {et~al.} 2018, \mnras, 475,
  648, \dodoi{10.1093/mnras/stx3112}

\bibitem[{{Planck Collaboration} {et~al.}(2020){Planck Collaboration},
  {Aghanim}, {Akrami}, {Ashdown}, {Aumont}, {Baccigalupi}, {Ballardini},
  {Banday}, {Barreiro}, {Bartolo}, {Basak}, {Battye}, {Benabed}, {Bernard},
  {Bersanelli}, {Bielewicz}, {Bock}, {Bond}, {Borrill}, {Bouchet}, {Boulanger},
  {Bucher}, {Burigana}, {Butler}, {Calabrese}, {Cardoso}, {Carron},
  {Challinor}, {Chiang}, {Chluba}, {Colombo}, {Combet}, {Contreras}, {Crill},
  {Cuttaia}, {de Bernardis}, {de Zotti}, {Delabrouille}, {Delouis}, {Di
  Valentino}, {Diego}, {Dor{\'e}}, {Douspis}, {Ducout}, {Dupac}, {Dusini},
  {Efstathiou}, {Elsner}, {En{\ss}lin}, {Eriksen}, {Fantaye}, {Farhang},
  {Fergusson}, {Fernandez-Cobos}, {Finelli}, {Forastieri}, {Frailis},
  {Fraisse}, {Franceschi}, {Frolov}, {Galeotta}, {Galli}, {Ganga},
  {G{\'e}nova-Santos}, {Gerbino}, {Ghosh}, {Gonz{\'a}lez-Nuevo}, {G{\'o}rski},
  {Gratton}, {Gruppuso}, {Gudmundsson}, {Hamann}, {Handley}, {Hansen},
  {Herranz}, {Hildebrandt}, {Hivon}, {Huang}, {Jaffe}, {Jones}, {Karakci},
  {Keih{\"a}nen}, {Keskitalo}, {Kiiveri}, {Kim}, {Kisner}, {Knox},
  {Krachmalnicoff}, {Kunz}, {Kurki-Suonio}, {Lagache}, {Lamarre}, {Lasenby},
  {Lattanzi}, {Lawrence}, {Le Jeune}, {Lemos}, {Lesgourgues}, {Levrier},
  {Lewis}, {Liguori}, {Lilje}, {Lilley}, {Lindholm}, {L{\'o}pez-Caniego},
  {Lubin}, {Ma}, {Mac{\'\i}as-P{\'e}rez}, {Maggio}, {Maino}, {Mandolesi},
  {Mangilli}, {Marcos-Caballero}, {Maris}, {Martin}, {Martinelli},
  {Mart{\'\i}nez-Gonz{\'a}lez}, {Matarrese}, {Mauri}, {McEwen}, {Meinhold},
  {Melchiorri}, {Mennella}, {Migliaccio}, {Millea}, {Mitra},
  {Miville-Desch{\^e}nes}, {Molinari}, {Montier}, {Morgante}, {Moss}, {Natoli},
  {N{\o}rgaard-Nielsen}, {Pagano}, {Paoletti}, {Partridge}, {Patanchon},
  {Peiris}, {Perrotta}, {Pettorino}, {Piacentini}, {Polastri}, {Polenta},
  {Puget}, {Rachen}, {Reinecke}, {Remazeilles}, {Renzi}, {Rocha}, {Rosset},
  {Roudier}, {Rubi{\~n}o-Mart{\'\i}n}, {Ruiz-Granados}, {Salvati}, {Sandri},
  {Savelainen}, {Scott}, {Shellard}, {Sirignano}, {Sirri}, {Spencer},
  {Sunyaev}, {Suur-Uski}, {Tauber}, {Tavagnacco}, {Tenti}, {Toffolatti},
  {Tomasi}, {Trombetti}, {Valenziano}, {Valiviita}, {Van Tent}, {Vibert},
  {Vielva}, {Villa}, {Vittorio}, {Wandelt}, {Wehus}, {White}, {White},
  {Zacchei}, \& {Zonca}}]{Planck18_20_A&A}
{Planck Collaboration}, {Aghanim}, N., {Akrami}, Y., {et~al.} 2020, \aap, 641,
  A6, \dodoi{10.1051/0004-6361/201833910}

\bibitem[{{Poon} {et~al.}(2023){Poon}, {Okabe}, {Fukazawa}, {Akino}, \&
  {Yang}}]{Poon_23_MNRAS}
{Poon}, H., {Okabe}, N., {Fukazawa}, Y., {Akino}, D., \& {Yang}, C. 2023,
  \mnras, 520, 6001, \dodoi{10.1093/mnras/stad514}

\bibitem[{{Popesso} {et~al.}(2024){Popesso}, {Biviano}, {Bulbul}, {Merloni},
  {Comparat}, {Clerc}, {Igo}, {Liu}, {Driver}, {Salvato}, {Brusa}, {Bahar},
  {Malavasi}, {Ghirardini}, {Robotham}, {Liske}, \&
  {Grandis}}]{Popesso_2024_MNRAS}
{Popesso}, P., {Biviano}, A., {Bulbul}, E., {et~al.} 2024, \mnras, 527, 895,
  \dodoi{10.1093/mnras/stad3253}

\bibitem[{{Predehl} {et~al.}(2021){Predehl}, {Andritschke}, {Arefiev},
  {Babyshkin}, {Batanov}, {Becker}, {B{\"o}hringer}, {Bogomolov}, {Boller},
  {Borm}, {Bornemann}, {Br{\"a}uninger}, {Br{\"u}ggen}, {Brunner}, {Brusa},
  {Bulbul}, {Buntov}, {Burwitz}, {Burkert}, {Clerc}, {Churazov}, {Coutinho},
  {Dauser}, {Dennerl}, {Doroshenko}, {Eder}, {Emberger}, {Eraerds},
  {Finoguenov}, {Freyberg}, {Friedrich}, {Friedrich}, {F{\"u}rmetz},
  {Georgakakis}, {Gilfanov}, {Granato}, {Grossberger}, {Gueguen}, {Gureev},
  {Haberl}, {H{\"a}lker}, {Hartner}, {Hasinger}, {Huber}, {Ji}, {Kienlin},
  {Kink}, {Korotkov}, {Kreykenbohm}, {Lamer}, {Lomakin}, {Lapshov}, {Liu},
  {Maitra}, {Meidinger}, {Menz}, {Merloni}, {Mernik}, {Mican}, {Mohr},
  {M{\"u}ller}, {Nandra}, {Nazarov}, {Pacaud}, {Pavlinsky}, {Perinati},
  {Pfeffermann}, {Pietschner}, {Ramos-Ceja}, {Rau}, {Reiffers}, {Reiprich},
  {Robrade}, {Salvato}, {Sanders}, {Santangelo}, {Sasaki}, {Scheuerle},
  {Schmid}, {Schmitt}, {Schwope}, {Shirshakov}, {Steinmetz}, {Stewart},
  {Str{\"u}der}, {Sunyaev}, {Tenzer}, {Tiedemann}, {Tr{\"u}mper}, {Voron},
  {Weber}, {Wilms}, \& {Yaroshenko}}]{Predehl_21_A&A}
{Predehl}, P., {Andritschke}, R., {Arefiev}, V., {et~al.} 2021, \aap, 647, A1,
  \dodoi{10.1051/0004-6361/202039313}

\bibitem[{{Rasmussen} {et~al.}(2006){Rasmussen}, {Ponman}, {Mulchaey}, {Miles},
  \& {Raychaudhury}}]{Rasmussen_2006_MNRAS}
{Rasmussen}, J., {Ponman}, T.~J., {Mulchaey}, J.~S., {Miles}, T.~A., \&
  {Raychaudhury}, S. 2006, \mnras, 373, 653,
  \dodoi{10.1111/j.1365-2966.2006.11023.x}

\bibitem[{{Salim} {et~al.}(2007){Salim}, {Rich}, {Charlot}, {Brinchmann},
  {Johnson}, {Schiminovich}, {Seibert}, {Mallery}, {Heckman}, {Forster},
  {Friedman}, {Martin}, {Morrissey}, {Neff}, {Small}, {Wyder}, {Bianchi},
  {Donas}, {Lee}, {Madore}, {Milliard}, {Szalay}, {Welsh}, \&
  {Yi}}]{Salim_2007_ApJS..173..267S}
{Salim}, S., {Rich}, R.~M., {Charlot}, S., {et~al.} 2007, \apjs, 173, 267,
  \dodoi{10.1086/519218}

\bibitem[{{Sanderson} {et~al.}(2003){Sanderson}, {Ponman}, {Finoguenov},
  {Lloyd-Davies}, \& {Markevitch}}]{Sanderson_2003_MNRAS}
{Sanderson}, A.~J.~R., {Ponman}, T.~J., {Finoguenov}, A., {Lloyd-Davies},
  E.~J., \& {Markevitch}, M. 2003, \mnras, 340, 989,
  \dodoi{10.1046/j.1365-8711.2003.06401.x}

\bibitem[{{Sarazin}(1988)}]{Sarazin_88_book}
{Sarazin}, C.~L. 1988, {X-ray emission from clusters of galaxies}, Cambridge
  Astrophysics Series (Cambridge: Cambridge Univ. Press)

\bibitem[{{Schellenberger} \& {Reiprich}(2017)}]{Schellenberger_2017_MNRAS}
{Schellenberger}, G., \& {Reiprich}, T.~H. 2017, \mnras, 469, 3738,
  \dodoi{10.1093/mnras/stx1022}

\bibitem[{{Shull} {et~al.}(2012){Shull}, {Smith}, \&
  {Danforth}}]{Shull_2012_ApJ...759...23S}
{Shull}, J.~M., {Smith}, B.~D., \& {Danforth}, C.~W. 2012, \apj, 759, 23,
  \dodoi{10.1088/0004-637X/759/1/23}

\bibitem[{{Sun}(2012)}]{Sun_2012_NJPH}
{Sun}, M. 2012, New Journal of Physics, 14, 045004,
  \dodoi{10.1088/1367-2630/14/4/045004}

\bibitem[{{Sun} {et~al.}(2009){Sun}, {Voit}, {Donahue}, {Jones}, {Forman}, \&
  {Vikhlinin}}]{Sun_09_APJ}
{Sun}, M., {Voit}, G.~M., {Donahue}, M., {et~al.} 2009, \apj, 693, 1142,
  \dodoi{10.1088/0004-637X/693/2/1142}

\bibitem[{{Truong} {et~al.}(2019){Truong}, {Rasia}, {Biffi}, {Mernier},
  {Werner}, {Gaspari}, {Borgani}, {Planelles}, {Fabjan}, \&
  {Murante}}]{Truong_2019_MNRAS}
{Truong}, N., {Rasia}, E., {Biffi}, V., {et~al.} 2019, \mnras, 484, 2896,
  \dodoi{10.1093/mnras/stz161}

\bibitem[{{Tully}(1987)}]{Tully_87_APJ}
{Tully}, R.~B. 1987, \apj, 321, 280, \dodoi{10.1086/165629}

\bibitem[{{Tuominen} {et~al.}(2021){Tuominen}, {Nevalainen}, {Tempel},
  {Kuutma}, {Wijers}, {Schaye}, {Hein{\"a}m{\"a}ki}, {Bonamente}, \&
  {Ganeshaiah Veena}}]{Tuominen_2021_A&A...646A.156T}
{Tuominen}, T., {Nevalainen}, J., {Tempel}, E., {et~al.} 2021, \aap, 646, A156,
  \dodoi{10.1051/0004-6361/202039221}

\bibitem[{{Watson} {et~al.}(2013){Watson}, {Iliev}, {D'Aloisio}, {Knebe},
  {Shapiro}, \& {Yepes}}]{Watson_2013_MNRAS}
{Watson}, W.~A., {Iliev}, I.~T., {D'Aloisio}, A., {et~al.} 2013, \mnras, 433,
  1230, \dodoi{10.1093/mnras/stt791}

\bibitem[{{Yang} {et~al.}(2005){Yang}, {Mo}, {van den Bosch}, \&
  {Jing}}]{Yang_2005_MNRAS}
{Yang}, X., {Mo}, H.~J., {van den Bosch}, F.~C., \& {Jing}, Y.~P. 2005, \mnras,
  356, 1293, \dodoi{10.1111/j.1365-2966.2005.08560.x}

\bibitem[{{Yang} {et~al.}(2007){Yang}, {Mo}, {van den Bosch}, {Pasquali}, {Li},
  \& {Barden}}]{Yang_2007_ApJ}
{Yang}, X., {Mo}, H.~J., {van den Bosch}, F.~C., {et~al.} 2007, \apj, 671, 153,
  \dodoi{10.1086/522027}

\bibitem[{{Yang} {et~al.}(2021){Yang}, {Xu}, {He}, {Gu}, {Katsianis}, {Meng},
  {Shi}, {Zou}, {Zhang}, {Liu}, {Wang}, {Dong}, {Lu}, {Li}, {Chen}, {Wang},
  {Mo}, {Fu}, {Guo}, {Leauthaud}, {Luo}, {Zhang}, \& {Zu}}]{Yang_2021_APJ}
{Yang}, X., {Xu}, H., {He}, M., {et~al.} 2021, \apj, 909, 143,
  \dodoi{10.3847/1538-4357/abddb2}

\bibitem[{{Zabludoff} \& {Mulchaey}(1998)}]{Zabludoff_1998_ApJ}
{Zabludoff}, A.~I., \& {Mulchaey}, J.~S. 1998, \apj, 496, 39,
  \dodoi{10.1086/305355}

\bibitem[{{Zhang} {et~al.}(2024){Zhang}, {Comparat}, {Ponti}, {Merloni},
  {Nandra}, {Haberl}, {Locatelli}, {Zhang}, {Sanders}, {Zheng}, {Liu},
  {Popesso}, {Liu}, {Truong}, {Pillepich}, {Predehl}, {Salvato}, {Shreeram},
  {Yeung}, \& {Ni}}]{Zhang_2024arXiv240117308Z}
{Zhang}, Y., {Comparat}, J., {Ponti}, G., {et~al.} 2024, arXiv e-prints,
  arXiv:2401.17308, \dodoi{10.48550/arXiv.2401.17308}

\bibitem[{{Zheng} {et~al.}(2023){Zheng}, {Yang}, {He}, {Shen}, {Li}, \&
  {Li}}]{Zheng_23_MNRAS}
{Zheng}, Y.-L., {Yang}, X., {He}, M., {et~al.} 2023, \mnras,
  \dodoi{10.1093/mnras/stad1684}

\end{thebibliography}

\end{document}